\documentclass[12pt]{article}
\usepackage{amssymb}
\usepackage[fleqn]{amsmath}
\usepackage[inline]{enumitem}
\usepackage{multirow}
\usepackage{xcolor}
\usepackage{verbatim}
\usepackage{graphicx}
\usepackage{float}
\usepackage{natbib}
\usepackage{color}
\usepackage[colorlinks,citecolor=blue,urlcolor=blue,filecolor=blue,backref=page]{hyperref}
\usepackage{url}
\usepackage{bbm}
\usepackage[colorlinks]{hyperref}
\usepackage[paperheight=12in,paperwidth=9in]{geometry}

\begin{document}

\begin{center}
	\Large \bf  Optimal Replacement Policy under Cumulative Damage Model and Strength Degradation with Applications 
\end{center}
\begin{center}
	\textbf{Phalguni Nanda$^{\ast}$ Prajamitra Bhuyan$^{\dag}$  Anup Dewanji$^{\S}$ }\\
	$^{\ast}$Department of Mathematics, National Institute of Technology, Rourkela\\
	$^{\dag}$Department of Mathematics, Imperial College London, London\\
	$^{\S}$Applied Statistics Unit, Indian Statistical Institute, Kolkata

\end{center}

\begin{abstract}
In many real-life scenarios, system failure depends on dynamic stress-strength interference, where strength degrades and stress accumulates concurrently over time.
In this paper, we consider the problem of finding an optimal replacement strategy that balances the cost of replacement with the cost of failure and results in a minimum expected cost per unit time under cumulative damage model with strength degradation. The existing recommendations are applicable only under restricted distributional assumptions and/or with fixed strength. As theoretical evaluation of the expected cost per unit time turns out to be very complicated, a simulation-based algorithm is proposed to evaluate the expected cost rate and find the optimal replacement strategy. The proposed method is easy to implement having wider domain of application. For illustration, the proposed method is applied to real case studies on mailbox and cell-phone battery experiments.
\end{abstract}

\noindent{\it Keywords: Convolution, Expected cost rate, Renewal process,  Simulation,  Grid search,  Simulated annealing}

\section{Introduction}\label{Intro}
The units or systems such as machines used in construction, chemical plants, power plants, heavy electrical and mechanical engineering, parts of vehicles, etc., are often subject to shocks in the course of their operation. These shocks may be assumed to appear at random points in time according to a point process and each shock causes some random amount of damage to the operating unit. The unit or system may fail at some sudden shock or it may withstand the shocks until the total damage caused by the shocks exceeds a critical level. The latter one is often encountered in practical situations and can be studied using a cumulative damage model. In this model, the damage caused in the form of crack growth, creep, fatigue, wear, etc., is accumulated until it becomes greater than a pre-specified threshold level. Some real life scenarios where this model turns out to be very helpful are discussed in the following. 

Crack in a vehicle axle caused by overload, jerk, etc., grows as long as it is above a certain depth and the axle breaks after that. \citet{Scarf1996} used a stochastic model under periodic inspection to study crack growth. Stochastic models were applied to study fatigue damage of materials by \citet{Sobczyk1987} and \citet{Sobczyk1992}. The electric power of a cell-phone battery, initially stored by chemical energy, is weakened by normal functionalities of a cell-phone and is subject to frequent calls leading to accumulated damage or energy loss \citep{Bhuyan2017b}.
Similarly, as a result of frequent updation of a database system, un-accessed data accumulates as garbage and the system collapsed as soon as it exceeds the tolerance level \citep[p-131]{Nakagawa2007}. As for example, the mailbox becomes full as a result of accumulation of emails over time and the account fails to receive any further email \citep{Bhuyan2017b}.
Keeping the unit or system functional until its failure may turn out to be cost-ineffective and lead to hazardous situations. If the axle of an automobile breaks in the course of its journey, then it may cost in terms of human lives, the goods it carries and extra money to repair. It creates a havoc among the users when servers in large systems such as banks, railways, online application programmes, etc., become unresponsive which often happens due to garbage created inside the database. Failure of units in nuclear power plants has proven its fatality in some events in the recent past. Hence, there is a need for preventive maintenance of the units before failure occurs \citep{Nakagawa2005}.

There has been ample research on the optimum replacement strategy assuming cumulative damage model with a constant strength or threshold level \citep[ch-3]{Nakagawa2007}. See also \citet{Taylor1975}, \citet{Zuckerman1977}, \citet{Chikte1977}, and \cite{Zhou2013} for work on replacement policies under similar damage accumulation models. All these works have assumed constant strength which may not be realistic in many practical situations. \citet{Zhou2016} proposed periodic preventive maintenance for leased equipment with continuous internal degradation and stochastic external shock damage. An operating unit is affected by human errors, material quality and operating conditions, etc., and the unit's capacity to withstand damage due to shocks may decrease as its operating time increases \citep{Satow2000}. Hence, the strength of a unit may reasonably be described by a deterministic curve which is decreasing in time. Recently, computation and estimation of reliability under such cumulative damage model has been considered by \citeauthor{Bhuyan2017a} (\citeyear{Bhuyan2017a}, \citeyear{Bhuyan2017b}).

In this article, we have discussed the replacement policies for the cumulative damage model having strength that is continuously non-increasing over time. In principle, we introduce a quantity called `expected cost per unit time' for each set of replacement (design) variables and minimize the same over the design variables to obtain the 'optimal' replacement policy.  Note that this expected cost per unit time depends on the distributions of the successive shock arrival times and also of the corresponding damages and the deterministic strength degradation curve in addition to the different cost components and the design variables (See Sections \ref{sec2} and \ref{sec3} for details). The computation of this expected cost per unit time is, however, often very challenging even for constant strength. Even if the distribution functions of both inter-arrival time between successive shocks and damage due to each shock possess closure property under convolution, the expression for the expected cost per unit time involves integrals and infinite sums, numerical evaluation of which is difficult. Complexity of computation increases if closed form expressions for the convolution of the associated distribution functions are not available and/or the strength is time dependent (See Section \ref{sec3}). In order to avoid such difficulty, \citet{Nakagawa1976b} and \citet{Endharta2014} assumed constant strength and independent and identically distributed (iid) Exponential distributions for the successive inter-arrival times (that is, the successive shock arrivals follow a homogeneous Poisson process) and damages so that the related convolutions follow the respective Gamma distributions. See also \citet{Satow2000}, however, for linearly decreasing strength curve. In this article, we propose a simulation based method for evaluation of the expected cost per unit time which provides flexibility in choosing the distribution functions for both inter-arrival time between successive shocks and damage due to each shock. Therefore, the domain of application of the proposed method is much wider.

In the next section, we discuss the preliminaries which include the notation and assumptions regarding the proposed modeling framework. In Section $\ref{sec3}$, we present the mathematical formulations for the basic replacement policies with different optimization criteria. Section $\ref{sec4}$ deals with the different computational methods and the issues therein. Some numerical results for different choices of the damage and inter-arrival time distributions, strength degradation for the unit, etc., are presented in Section $\ref{sec5}$. In Section $\ref{sec6}$, we consider some generalizations of the damage distribution and present some numerical results in those cases. We illustrate the proposed method using real case studies in Section \ref{case}. Finally, we conclude the paper with some remarks in Section $\ref{sec7}$.

\section{Preliminaries}\label{sec2}
We assume that the operating unit starts working at time $0$ and its initial damage level is $0$. As time progresses, it is subject to shocks and suffers from some amount of damage due to each shock. These damages caused by the successive shocks are accumulated over time. Let $N(t)$ represent the number of shocks by time $t$. It is assumed that the shocks arrive according to a renewal process. Let $X_{1}, X_{2}, \ldots$ be the sequence of independent and identically distributed random variables which denote the inter-arrival times between successive shocks having the common distribution function $F(\cdot)$. Then $S_{j} = \sum_{i=1}^{j} X_{i},~j \geq 1$, represents the arrival time of the $j$th shock and has the distribution function $F^{(j)}(\cdot)$, where $F^{(j)}(\cdot)$ is the $j$-fold convolution of $F(\cdot)$ with itself. The successive damages $W_{1}, W_{2},\ldots$ are assumed to be independent and identically distributed and also independent of the shock arrival process $N(t)$ (that is, the $X_i$'s). Let $W_{j},~j\geq 1$, have a common distribution function $G(\cdot)$. Then the total damage at the $j$th shock will have the distribution function $G^{(j)}(\cdot)$, the $j$-fold convolution of $G(\cdot)$ with itself.

The strength of the unit is described by $K(t)$ which is continuous and decreasing in time $t$. Note that, under the present stress-strength interface, there are two different types of failure modes, either due to strength degradation at or below the existing level of accumulated stress, or due to arrival of a shock resulting in the increased stress
exceeding or equaling the strength at that time (See \citeauthor{Bhuyan2017a}, \citeyear{Bhuyan2017a}). Then a unit fails when its strength reduces to zero even if no shock arrives by that time. One needs to consider corrective replacement of the unit with a new one immediately after failure. According to the existing basic replacement policies, the unit is preventively replaced before failure at a planned time $T$, or a shock number $N$, or a damage level $Z$, whichever occurs first; otherwise it is replaced at failure (corrective replacement). In our work, we have adopted the basic replacement policies with an additional condition $Z \leq K(T)$ so that the damage level $Z$ has some relevance in deciding the replacement policies. If the total damage at the $N$th shock exceeds the pre-specified damage level $Z$, or the strength at that time of shock arrival, then it is assumed that the replacement of the unit is due to damage, or failure, as is the case, instead of the shock number $N$. This assumption is reasonable if both the replacement costs, due to damage $Z$ and due to failure, are higher than that due to shock number $N$, in order to safeguard the worse situation. Similarly, if the total damage at the $N$th shock exceeds both the damage level $Z$ and the strength at that time of shock arrival, we assume that the replacement is due to the failure, since that is presumably the most expensive of the three. 

Let us denote the probabilities that the unit is replaced at scheduled time $T$, shock number $N$, damage level $Z$ and at failure, by $p_T$, $p_N$, $p_Z$ and $p_K$, respectively. We assume that all replacements are instantaneous. There is cost associated with each replacement with the cost of corrective replacement being higher than those of the preventive replacement. If $c_{T}$, $c_{N}$, $c_{Z}$, $c_{K}$ are the costs incurred from replacement at time $T$, shock number $N$, damage level $Z$ and at failure, respectively, then $c_{K}$ is higher than each of $c_{T}$, $c_{N}$ and $c_{Z}$. The expected cost for replacement can be obtained as a function of the design variables $T$, $N$ and $Z$, denoted by $\tilde{C}(T, N, Z)$, which upon division by the mean time to replacement gives the expected replacement cost per unit time, termed as the `expected cost rate' for brevity. The expected cost rate as defined above is known as `long run mean cost' in the context of renewal process theory which requires the process to be regenerative, or renewed, after each replacement, preventive or corrective. In the stress-strength interference leading to the cumulative damage model, this regenerative or renewal property does not hold in general, since the shock arrivals may not start anew after each replacement. Nevertheless, if the shock arrival is modeled by a homogeneous Poisson process (HPP), it behaves like starting anew at each replacement time due to the memoryless property; therefore, the whole stress-strength interference starts anew at each replacement time ensuring the renewal property. For non-HPP
shock arrivals, one can think of three alternatives. First, depending on the situation, the shock arrivals may be linked with the functioning of the device, like a particular type of stressful uses (See both the real examples in \citet{Bhuyan2017b}), in which case the renewal property
at each replacement time is a clear consequence. Secondly, one may be interested in the case of only the first replacement in which case the expected cost rate may be interpreted as the average cost per unit time until the first replacement \citep{Rafiee}. Thirdly, the system can only undergo a limited number of replacements in practice after which the system becomes outdated. Noting that only the first shock arrival time after a replacement has a different (in fact, residual life) distribution, the expected cost rate may be taken as an approximation to the `long run mean cost' under this non-renewal point process and, hence, a reasonable objective function to minimize. Notwithstanding this difficulty associated with the definition of the expected cost rate, we henceforth consider this as the objective function in view of the above discussion. \citet{Sheu} considered a similar objective function, termed as `long term expected cost per unit time', for optimization of two replacement policies in $k$-out-of-$n$ systems. See also \citet{Lee}, \citet{Zhao} and \citet{Ery} among others for  consideration of similar objective function to obtain optimal replacement policies in different contexts and modeling scenarios.

\section{Optimal Replacement Policies}\label{sec3}
As described in the previous section, a preventive replacement is to be carried out at a planned time $T$, or at a shock number $N$, or at a damage level $Z$, whichever occurs first. As in \citet{Satow2000}, we first consider these three design variables $T$, $N$ and $Z$ one at a time and consider the corresponding expected cost rates as the objective function to minimize. However, the expressions for the expected cost rates are different because of the time-dependent strength degradation. Thereafter, we deal with all these three variables simultaneously. For this purpose, we derive the expected cost rates for replacement separately as a function of $T$, $N$ and $Z$ and then all taken together. In the following, for the ease of understanding, we present simply the respective expressions for the expected cost rates with some reference to the materials in the Appendix, where details of the derivations are presented.

We first discuss the preventive replacement of the unit only at a planned time $T$. The unit is replaced either at $T$ or at failure, whichever occurs first. There is no replacement at the $N$th shock or the cumulative damage reaching $Z$. As discussed in the previous section, we assume that the replacement is corrective rather than preventive, if failure happens at time $T$. 
Then, the expected cost rate $C_{1}(T)$, when the unit is replaced either at $T$ or at failure, can be obtained, dividing $(\ref{eq3.2})$ by $(\ref{eq3.3})$, as
\begin{equation}\label{eq3.4}
C_{1}(T) = \frac{c_{K} - \left(c_{K} - c_{T}\right) \left[\bar{F}(t) + \sum_{j=1}^{\infty} \left[ F^{(j)}(T) - F^{(j+1)}(T) \right]G^{(j)}(K(T)) \right]}{\int_{0}^{T}\bar{F}(t)dt + \sum_{j=1}^{\infty} \int_{0}^{T} \left[F^{(j)}(t) - F^{(j+1)}(t)\right]G^{(j)}(K(t)) dt}.
\end{equation}

When $K(T)=K$, for the case of constant strength, Eq. \eqref{eq3.4} simplifies to that of Eq. (3.11) of \citet[p-42]{Nakagawa2007}.

Next we consider the case when the operating unit is replaced either at the planned shock number $N$ or at failure, whichever occurs first. There is no replacement at a planned time $T$ or due to reaching a damage level $Z$. As discussed before, we assume that the replacement is corrective rather than preventive, if failure happens at the arrival of the $N$th shock. The expected cost rate $C_{2}(N)$ for replacement is, dividing (\ref{eq3.6}) by (\ref{eq3.7}), given by
\begin{equation}\label{eq3.8}
C_{2}(N) = \frac{c_{K} - \left(c_{K} - c_{N}\right)\int_{0}^{\infty} G^{(N)}(K(s)) dF^{(N)}(s)}{\mu_{F} + \sum_{j=1}^{N-1} \int_{0}^{\infty} \left[F^{(j)} - F^{(j+1)}\right]G^{(j)}(K(t)) dt}.
\end{equation}

When $K(s)=K$, for the case of constant strength, Eq. \eqref{eq3.8} simplifies to that of Eq. (3.20) of \citet[p-44]{Nakagawa2007}.

Now we consider the problem of replacement at a planned cumulative damage level $Z$ or at failure, whichever occurs first. There is no replacement at the planned time $T$ or at the $N$th shock.  
Here, the expected cost rate for replacement, denoted by $C_{3}(Z)$, is obtained, dividing (\ref{eq3.9}) by (\ref{eq3.12}), as
\begin{eqnarray}\label{eq3.13}
C_{3}(Z) &=& c_{K} - \left(c_{K} - c_{Z}\right)\Big[ \int_{0}^{T_{0}} \left[G(K(s)) - G(Z)\right] dF(s) \nonumber \\ && + \sum_{j=1}^{\infty} \int_{0}^{T_{0}} \int_{0}^{Z} \left[ G(K(s) - x) - G(Z - x) \right] dG^{(j)}(x) dF^{(j+1)}(s) \Big]\nonumber \\ && \Bigg/ \Big[ \mu_{F} + \sum_{j=1}^{\infty} \int_{0}^{\infty} \left[F^{(j)}(t) - F^{(j+1)}(t)\right]G^{(j)}(\tilde{K}(t)) dt \Big]. 
\end{eqnarray}

When $K(s)=K$, for the case of constant strength, Eq. \eqref{eq3.13} simplifies to that of Eq. (3.24) of \citet[p-45]{Nakagawa2007}.

Finally, we consider preventive replacement under simultaneous consideration of $T$, $N$ and $Z$. Replacement of the unit takes place at a planned time $T$, shock number $N$, at a damage level $Z$,  or at failure, whichever occurs first. As discussed before, if the cumulative damage at the $N$th shock exceeds $Z$ as well as the strength at that time, we assume that the replacement is corrective, since that is more expensive compared to preventive replacement. The expected cost rate of replacement  in this case, denoted by $C(T, N, Z)$, is obtained, dividing (\ref{eq3.18}) by (\ref{eq3.19}), as 
\begin{eqnarray} \label{eq3.20}
C(T, N, Z)&=& \Bigg[c_{K} - \left(c_{K} - c_{T}\right)\Big[\bar{F}(T) + \sum_{j=1}^{N-1} G^{(j)}(Z)\left\lbrace F^{(j)}(T) - F^{(j+1)}(T) \right\rbrace \Big]  \nonumber\\
&& -  \left(c_{K} - c_{N}\right) F^{(N)}(T) G^{(N)}(Z) - \left(c_{K} - c_{Z}\right) \Big\{\int_{0}^{T} \left[G(K(s)-G(z))\right]dF(s) \nonumber \\ && + \sum_{j=1}^{N-1} \int_{0}^{T} \int_{0}^{Z} \left[ G(K(s) - x) - G(Z - x) \right] dG^{(j)}(x) dF^{(j+1)}(s)\Big\} \Bigg] \nonumber \\ && \big/ \Bigg[\int_{0}^{T}\bar{F}(t)dt  + \sum_{j=1}^{N-1} G^{(j)}(Z) \int_{0}^{T} \left[F^{(j)}(t) - F^{(j+1)}(t)\right] dt\Bigg]. 
\end{eqnarray}

When $K(s)=K$, for the case of constant strength, Eq. \eqref{eq3.20} simplifies to that of Eq. (3.8) of \citet[p-42]{Nakagawa2007}.

Our objective is to find the optimum choices of $T$, $N$ and $Z$ which minimize the respective expected cost rates, under the four different design considerations described above, in the corresponding design space. Theoretically optimizing the expected cost rates leads to complicated expressions and requires imposing more conditions which are practically less important. Moreover, no analytical solution of the optimum replacement policy is available, even if the damage distribution and that of the inter-arrival times possess closure property under convolution like, for example, the exponential distribution. See \citet[Ch-3]{Nakagawa2007} for details. Thus, there is a need to go for numerical investigation for finding an approximation of the optimum replacement policy, denoted by $\hat{T}$, $\hat{N}$, and $\hat{Z}$, respectively. Henceforth, we refer to this approximation as the `optimum replacement policy' although this is only approximately optimal. The methods and the issues associated with this investigation are discussed in the following section.

\section{Computational Issues}\label{sec4}
As remarked in Section \ref{Intro}, the computation of the expected cost rates, given by Eq. \eqref{eq3.4}, \eqref{eq3.8}, \eqref{eq3.13}, and \eqref{eq3.20}, is extremely challenging. The expressions for expected cost rates involve infinite sums or infinite integrals, or both in some cases. Evaluation of the integrals can be carried out using numerical integration. Any standard software package equipped with numerical integration (e.g.- \textit{integrate} in the package R) can be used for this purpose. The details of the algorithm and its precision are discussed in \citet{Piessens1983}. On the other hand, the infinite sums can be approximated by taking large number, say 10000, of terms and ignoring the terms after that. Evaluation of the expected cost rate using this approach is computationally challenging but feasible if both of the inter-arrival time between successive shocks and damage due to each shock follow exponential distributions. However, as mentioned before, analytical solution of the optimum replacement policy is not available even for this simple scenario. The complexity of computation increases if the distribution functions do not have closure property under convolution (e.g., Weibull, Log-normal, etc.). To address this difficulty in numerically obtaining the expected cost rates in other situations, we resort to a method of simulation, as described below, to obtain the expected cost rates approximately.

In this method, the whole process of shock arrivals and accumulation of damages as against the degradation  of strength is virtually created. For a fixed $T$, $N$ and $Z$, the proposed algorithm gives as output one realization each for the time to replacement $T_{R}$ and a variable $I_{R}$ indicating whether the replacement is due to failure or due to one of $N,\ T$ and $Z$ taking values 0, 1, 2 and 3, respectively. The mean time to replacement and the probabilities of replacement can be estimated by simulating a large number, say $10000$, of  realizations of $T_{R}$ and $I_{R}$. The algorithm for simulating a realization for each of $T_{R}$ and $I_{R}$ is given below :
\begin{enumerate}
	\item[] $X_{0} = 0$, $W_{0} = 0$; For $i = 1, 2, \ldots$,
	\item[Step 1.] Simulate $X_{i} \sim F(\cdot)$ and $W_{i} \sim G(\cdot)$;
	\item[Step 2.] Calculate $S_{i} = \sum_{j=0}^{i} X_{j}$ and $L_{i} = \sum_{j=0}^{i} W_{j}$;
	\item[Step 3.] If $L_{i} < \tilde{K}(S_{i})$, then next $i$ (i.e. repeat Step 1 and Step 2);\\
	else, if $L_{i-1} \leq \tilde{K}(S_{i}) < L_{i}$, then set $\tilde{T} = S_{i}$;\\
	else, find $\tilde{T} = t$ by solving $\tilde{K}(t) = L_{i-1}$;
	\item[Step 4.] If $N(\tilde{T}) > N$, then $T_{R} = \min \left\lbrace T, S_{N} \right\rbrace$ and $I_{R} = 1\mathbbm{1}(T_{R}=S_{N})+2\mathbbm{1}(T_{R}=T)$;\\
	else, if $\tilde{T} < T_{0}$ and $L_{i} < K(S_{i})$, then $T_{R} = \min \left\lbrace \tilde{T}, T \right\rbrace $ and $I_{R} = 2\mathbbm{1}(T_{R}=T)+3\mathbbm{1}(T_{R}=\tilde{T})$;\\
	else, if $\tilde{T} < T$, then $T_{R} = \tilde{T}$ and $I_{R} = 0$;\\
	else, $T_{R} = T$ and $I_{R} = 2$.
\end{enumerate}
Now, the expected cost rate $C(T,N,Z)$ given in \eqref{eq3.20} is approximated as $\hat{C}(T,N,Z)=\frac{c_{T}\bar{I}_{T} + c_{Z}\bar{I}_{Z}+ c_{N}\bar{I}_{N}+ c_{K}(1-\bar{I}_{T}-\bar{I}_{Z}-\bar{I}_{N})  }{\bar{T}_{R}}$, where $\bar{I}_{T}$, $\bar{I}_{Z}$, $\bar{I}_{N}$ and $\bar{T}_{R}$ are mean of $I_{T}=\mathbbm{1}(I_{R}=2)$, $I_{Z}=\mathbbm{1}(I_{R}=3)$, $I_{N}=\mathbbm{1}(I_{R}=1)$ and $T_{R}$, respectively, based on $10000$ simulated observations. The proposed algorithm evaluates the expected cost rate as a function of $T$, $N$ and $Z$ which can then be minimized for finding the optimal values of $T$, $N$ and $Z$ for replacement. The optimal replacement, while considering one of $T$, $N$ and $Z$ (See Section \ref{sec3}), can be determined by minimizing $\hat{C_{1}}(T)=\hat{C}(T, \infty,\infty)$, $\hat{C_{2}}(N)=\hat{C}(\infty,N,\infty)$, and $\hat{C_{3}}(Z)=\hat{C}(\infty,\infty,Z)$, respectively.
The minimum expected cost rate can be obtained by using the methods of grid search, simulated annealing, etc. Note that the approximated expected cost rate obtained by the aforementioned simulation algorithm may consist of some local minimums. The method of simulated annealing has been implemented to escape a local minimum with certain probability in order to search for the global minimum. Interested readers can see \citet{Kirkpatrick1983} and \citet{Dowsland1995} for more details on simulated annealing. Since the number of design parameters is small, sequential grid search also preforms efficiently.
It is important to note that the domain of application of the proposed simulation method is much wider providing flexibility in choosing both the distribution functions for inter-arrival time between successive shocks and damage due to each shock. 

\section{Numerical Results}\label{sec5}
The computations have been done under different distributional assumptions with several sets of values for the associated parameters, different strength degradation and the cost incurred from replacement at failure. In all of the computations, the costs incurred from preventive replacements at $T$, $N$ or $Z$ are assumed to be $1$, i.e. $c_{T} = c_{N} = c_{Z} = 1$.
The inter-arrival time between successive shocks has been assumed to follow 
\begin{enumerate*}[label = (\itshape\roman*\upshape)] 
	\item Exponential distribution with mean $1 / \lambda$, denoted by $Exp(\lambda)$, and \item Log-normal distribution with Normal parameters $\mu$ and $\sigma$, denoted by $LN(\mu, \sigma)$, with mean being $\exp\left(\mu+\frac{1}{2}\sigma^{2}\right)$.
\end{enumerate*} The distribution functions for the damage caused by each shock has been assumed to be either
\begin{enumerate*}[label = (\itshape\roman*\upshape)]
	\item Exponential with mean $1 / \mu$, denoted by $Exp(\mu)$, or \item Weibull with scale parameter $\alpha$ and shape parameter $\beta$, denoted by $Wei(\alpha, \beta)$, with mean damage being $\alpha\Gamma\left(1+\frac{1}{\beta}\right)$.
\end{enumerate*} The strength degradation curve $K(t)$ is assumed to be exponential, linear or constant over time. 

In Table \ref{tab1}, we present the optimum values $\hat{T}$, $\hat{N}$ and $\hat{Z}$ which minimize the approximate expected cost rates $\hat{C}_{1}(T)$, $\hat{C}_{2}(N)$ and $\hat{C}_{3}(Z)$, respectively, along with the corresponding minimum expected cost rates. Then, in Table \ref{tab2}, we present the optimum values $\hat{T}$, $\hat{N}$ and $\hat{Z}$ by minimizing  the approximate expected cost rate $\hat{C}(T, N, Z)$ as a function of $T$, $N$ and $Z$, along with the corresponding minimum expected cost rate. In Table \ref{uneqaul}, a different set of cost components ($c_T=0.5, c_N=1.5, c_Z=1.0$ and $c_K=6$) is considered for the optimum values  $\hat{T}, \hat{N}$ and $\hat{Z}$, corresponding to simultaneous optimization as in Table \ref{tab2}, to study the impact of differential cost component.  When both $F$ and $G$ are Exponential, then one can compute the expected cost rates, given by Eq. \eqref{eq3.4}, \eqref{eq3.8}, \eqref{eq3.13}, and \eqref{eq3.20}, directly (as remarked at the beginning of Section \ref{sec4}) and the optimization results are obtained by implementing the grid search method. This method, used only when both $F$ and $G$ are Exponential, is termed as `Direct: GS' in the tables (GS meaning `grid search'). Otherwise, when the expected cost rates are approximated by simulation, optimization results are obtained by using the grid search and/or the simulated annealing algorithm, termed as `Approx: GS' and `Approx: SA' (SA meaning `simulated  annealing'), respectively, in the tables. These latter two methods have sometime been used for comparison even when both $F$ and $G$ are Exponential. It is clearly seen that, when both $F$ and $G$ are Exponential,  the results based on the approximate expected cost rates using the simulation method are similar to those obtained by the direct method (See Tables \ref{tab1}-\ref{uneqaul}).
As expected, one can observe that the optimal values $\hat{T}$, $\hat{N}$ and $\hat{Z}$ decrease as cost of corrective replacement $c_{K}$ increases (See Table \ref{tab1}). Also, as expected, the optimal values of $T$, $N$ and $Z$ in Table \ref{tab2} are larger compared to those in Table \ref{tab1}, since the condition of replacement in Table \ref{tab2} is more stringent (any of the design variables $T$ or $N$ or $Z$ exceeds the respective threshold). Interestingly, the minimum expected cost rate is smaller for the simultaneous optimization of $T$, $N$ and $Z$ compared to those of individual cases, as expected, since the domain of minimization is smaller in the individual cases. Note that, in Table \ref{uneqaul}, $\hat{N}$ and $\hat{Z}$ are larger compared to those in Table \ref{tab2}, as expected, since the costs $c_N$ and $c_Z$ are higher. So, it conforms with the natural trend that, if a cost component is higher, the corresponding threshold tends to be higher to safeguard against that cost. 

\begin{center}
	\begin{table}[h!]
		\centering
		\caption{Optimal $\hat{T}$, $\hat{N}$, $\hat{Z}$ and the corresponding minimal expected cost rates $C_{1}(\hat{T})$, $C_{2}(\hat{N})$ and $C_{3}(\hat{Z})$ with $c_{T} = c_{N} = c_{Z} = 1$. Means of the relevant distributions are given in parentheses.}  
		\label{tab1}
		\resizebox{\columnwidth}{6 cm}{
			\begin{tabular}{c | c | c | c | c | c c | c c | c c}
				\hline
				$K(t)$ & $F$ & $G$ & $c_{K}$ & Method & $\hat{T}$ & $C_{1}(\hat{T})$ & $\hat{N}$ & $C_{2}(\hat{N})$ & $\hat{Z}$ & $C_{3}(\hat{Z})$\\
				\hline
				$100\exp(-0.1 t)$ & $Exp(0.4)$ & $Exp(4)$ & 2 & Approx: GS & 29.42 & 0.036 & 10 & 0.044 &2.45 &0.046\\
				&  &  &   & Direct: GS & $29.34$ & $0.035$ & $10$ & $0.043$ & $2.51$ & $0.046$\\
				& $(2.5)$ & $(0.25)$ & $4$  & Approx: GS & $28.23$ & $0.037$ & $9$ & $0.049$ & $1.84$ & $0.056$\\
				&  &  &   & Direct: GS & $28.06$ & $0.037$ & $9$ & $0.049$ & $1.92$ & $0.056$\\
				&  &  & 6 & Approx: GS & 27.29 &0.037 &9 &0.053 & 1.70 &0.060\\
				&  &  &   & Direct: GS & $27.57$ & $0.037$ & $9$ & $0.054$ & $1.72$ & $0.061$\\
				\hline
				$\max\left\lbrace50 - t, 0\right\rbrace$ & $Exp(0.5)$ & $Exp(0.5)$ & 2 & Grid Search & 20.53 & 0.057 & 10 & 0.058 & 18.53 & 0.057\\
				&  &  &   & Direct: GS & $20.48$ & $0.058$ & $10$ & $0.057$ & $18.47$ & $0.058$\\
				& (2) & (2)  & 4 & Approx: GS & 17.19 & 0.067 & 9 & 0.066 & 15.07 & 0.066\\
				&  &  &   & Direct: GS & $17.33$ & $0.067$ & $9$ & $0.066$ & $15.33$ & $0.066$\\
				&  & & 6 & Approx: GS & 16.03 & 0.071 & 8 & 0.070 & 14.39 & 0.070\\
				&  &  &   & Direct: GS & $16.15$ & $0.071$ & $8$ & $0.070$ & $14.15$ & $0.071$\\
				\hline
				$10$ & $Exp(0.5)$ & $Exp(1)$ & $2$ & Approx: GS & $20.21$ & $0.084$ & 9 & $0.078$ & $7.95$ & $0.063$\\
				&  &  &   & Direct: GS & $20.25$ & $0.084$ & $9$ & $0.078$ & $7.93$ & $0.063$\\
				& (2)  & (1) & $4$ & Approx: GS & $12.76$ & $0.119$ & $6$ & $0.100$ & $6.92$ & $0.071$\\
				&  &  &   & Direct: GS & $12.76$ & $0.119$ & $6$ & $0.101$ & $6.96$ & $0.072$\\
				&  &  & 6 & Approx: GS & 10.83 & 0.139 & 6 & 0.113 & 6.57 & 0.078 \\	
					&  &  &   & Direct: GS & $10.64$ & $0.139$ & $6$ & $0.112$ & $6.51$ & $0.077$\\
				\hline
				\hline
				$150\exp(-0.05t)$ & $LN(2, 1)$ & $Wei(10,15)$ & 2 & Approx: GS & $26.09$ & $0.042$ & $3$ & $0.046$ & $21.13$ & $0.046$\\
				& $(12.18)$ & $(9.66)$ & $4$ & Approx: GS & $21.96$ & $0.047$ & $2$ & $0.062$ & $13.16$ & $0.062$\\
				&  &  & 6 & Approx: GS & 21.85 & 0.049 &2 & 0.074 & 13.90 &0.074\\
				\hline
				$\max\left\lbrace60 - t, 0\right\rbrace$ & $LN(1, 1)$ & $Wei(10, 5)$ & $2$ & Approx: GS & $15.47$ & $0.089$ & $4$  & $0.073$ & $30.25$ & $0.072$\\				
				& $(4.48)$ & $(9.18)$ & $4$ & Approx: GS & $11.56$ & $0.108$ & $3$ & $0.086$ & $24.74$ & $0.086$\\
				&  &  &6 & Approx: GS &9.72 & 0.120 &3 &0.095 &22.59 &0.095\\
				\hline		
				$50$ & $LN(2, 1)$ & $Wei(10,15)$ & $2$ & Approx: GS & $74.72$ & $0.028$ & $5$ & $0.019$ & $39.63$ & $0.018$\\			
				& $(12.18)$ & $(9.66)$ & $4$ & Approx: GS & $35.18$ & $0.038$ & $4$  & $0.021$ & $39.30$ & $0.018$\\
				&  &  &6 & Approx: GS &29.84 & 0.043& 4 & 0.021& 37.71&0.018\\
				\hline
		\end{tabular}}
	\end{table}
\end{center}

\begin{center}
	\begin{table}[H]
		\centering
		\caption{Optimal $\hat{T}$, $\hat{N}$, $\hat{Z}$ and the corresponding minimal expected cost rate $C(\hat{T}, \hat{N}, \hat{Z})$ with $c_{T} = c_{N} = c_{Z} = 1$. Means of the relevant distributions are given in parentheses. }
		\label{tab2}
		\resizebox{\columnwidth}{3 cm}{
			\begin{tabular}{c | c | c | c | c | c c c c}
				\hline
				$K(t)$ & $F$ & $G$ & $c_{K}$ & Method & $\hat{T}$ & $\hat{N}$ & $\hat{Z}$ & $C(\hat{T}, \hat{N}, \hat{Z})$\\
				\hline
				$100\exp(-0.1t)$ & $Exp(0.4)$ & $Exp(4)$ & $4$ & Approx: GS & $31.40$ & $19$ & $4.21$ & $0.033$\\
				& $(2.5)$ & $(0.25)$ &  & Approx: SA &  $30.99$ & $18$ & $4.20$ & $0.034$\\
				&  &  &  & Direct: GS &  $31.20$ & $19$ & $4.20$ & $0.034$\\
				\hline
				$\max\left\lbrace 50 - t, 0 \right\rbrace$ & $Exp(0.5)$ & $Exp(0.5)$ & $6$ & Approx: GS & $25.01$ & $13$ & $20.40$ & $0.051$\\
				& $(2)$ & $(2)$ & & Approx: SA & $25.03$ & $13$ & $19.91$ & $0.051$\\
				&  &  &  & Direct: GS &  $24.20$ & $13$ & $21.50$ & $0.052$\\
				\hline
				\hline
				$150\exp(-0.05t)$ & $LN(2, 1)$ & $Wei(10,15)$ & $2$ & Approx: GS & $35.02$ & $4$ & $25.87$ & $0.036$\\
				& $(12.18)$ & $(9.66)$ & & Approx: SA &  $34.12$ & $4$ & $24.69$ & $0.037$\\
				\hline
				$\max\left\lbrace 60 - t, 0 \right\rbrace$ & $LN(1, 1)$ & $Wei(10,5)$ & $4$ & Approx: GS & $30.41$ & $4$ & $23.74$ & $0.067$\\
				& $(4.48)$ & $(9.18)$ & & Approx: SA &  $30.72$ & $4$ & $23.06$ & $0.067$\\
				\hline
		\end{tabular}}
	\end{table}
\end{center}

\begin{center}
	\begin{table}[H]
		\centering
		\caption{Optimal $\hat{T}$, $\hat{N}$, $\hat{Z}$ and the corresponding minimal expected cost rate $C(\hat{T}, \hat{N}, \hat{Z})$ with $c_{K}=6, c_{T} =0.5 , c_{N} =1.5 , c_{Z} =1 $. Means of the relevant distributions are given in parentheses. }
		\label{uneqaul}
		\resizebox{\columnwidth}{3 cm}{
			\begin{tabular}{c | c | c | c | c  c c c c}
				\hline
				$K(t)$ & $F$ & $G$ &  Method & $\hat{T}$ & $\hat{N}$ & $\hat{Z}$ & $C(\hat{T}, \hat{N}, \hat{Z})$\\
				\hline
				$100\exp(-0.1t)$ & $Exp(0.4)$ & $Exp(4)$ & Approx: GS & $28.57$ & $26$ & $5.41$ & $0.018$\\
				& $(2.5)$ & $(0.25)$ & Approx: SA &  $28.89$ & $26$ & $5.38$ & $0.018$\\
				&  &  &  Direct: GS &  $28.66$ & $26$ & $5.42$ & $0.018$\\
				\hline
				$\max\left\lbrace 50 - t, 0 \right\rbrace$ & $Exp(0.5)$ & $Exp(0.5)$ & Approx: GS & $18.41$ & $21$ & $28.47$ & $0.033$\\
				& $(2)$ & $(2)$ & Approx: SA & $17.23$ & $21$ & $30.13$ & $0.033$\\
				&  &  &  Direct: GS &  $18.73$ & $21$ & $28.91$ & $0.033$\\
				\hline
				\hline
				$150\exp(-0.05t)$ & $LN(2, 1)$ & $Wei(10,15)$ & Approx: GS & $22.72$ & $8$ & $45.10$ & $0.024$\\
				& $(12.18)$ & $(9.66)$ & Approx: SA &  $22.05$ & $8$ & $44.29$ & $0.024$\\
				\hline
				$\max\left\lbrace 60 - t, 0 \right\rbrace$ & $LN(1, 1)$ & $Wei(10,5)$  & Approx: GS & $13.41$ & $7$ & $37.01$ & $0.055$\\
				& $(4.48)$ & $(9.18)$ & Approx: SA &  $13.93$ & $7$ & $37.32$ & $0.055$\\
				\hline
		\end{tabular}}
	\end{table}
\end{center}

\section{Some Generalizations }\label{sec6}
In this section, we consider some generalizations in the assumption related to the successive damage distributions which may be more realistic in some situations. The damages due to shocks may be either dependent or independent but not identically distributed. As we move on to these generalized scenarios, the computational difficulty associated with the direct method also increases. In such situations, the simulation method turns out to be more effective. The algorithm for simulation remains similar to that described in Section $\ref{sec4}$ except for the damage distributions for simulating the $W_i$'s which change accordingly. The optimal values of $T$, $N$ and $Z$ and the corresponding minimum expected cost rates are evaluated in the same manner.

\subsection{Independent but Non-iid Damage Distributions}
Here we assume that the damages caused by the successive shocks may be independent but not identically distributed. For instance, there may be situations where the successive shocks cause damages which are stochastically larger than those due to the preceding ones. Note that when the damages $X_{1}, X_{2},\ldots$ are independent but not identically distributed, then $Y = \sum_{i = 0}^{n} X_{i}$ 
under these assumptions may not fall into any known class of distributions. As mentioned before, there are several difficulties in evaluating the expected cost rates directly since the expressions are not in a closed form. Interestingly, the algorithm for the simulation method remains the same except that the successive damages are now generated from the non-identical distributions and can be easily implemented.


The optimal values of $T$, $N$ and $Z$ and the corresponding minimum values of expected cost rates $C_{1}(T)$, $C_{2}(N)$ and $C_{3}(Z)$ under different distributional assumptions are presented in Table $\ref{tab3}$. The shocks are assumed to arrive according to a renewal process, i.e. the inter-arrival time between successive shocks are iid with a common distribution function $F(\cdot)$. We have chosen the inter-arrival time distribution to be 
\begin{enumerate*}[label = (\itshape\roman*\upshape)] 
	\item Exponential distribution with mean $1 / \lambda$, denoted by $Exp(\lambda)$, \item Log-normal distribution with Normal parameters $\mu$ and $\sigma$, denoted by $LN(\mu, \sigma)$.
\end{enumerate*} Unlike the case of iid damages, here it is assumed that the damage due to $i$th shock has a distribution function $G_{i}(\cdot)$. The choices for $G_{i}(\cdot)$ are
\begin{enumerate*}[label = (\itshape\roman*\upshape)]
	\item Gamma with scale parameter $\theta_{i}$ and shape parameter $\delta$, denoted by $Ga(\theta_{i}, \delta)$, with mean being $\delta \theta_{i}$ or \item Weibull with scale parameter $\alpha_{i}$ and shape parameter $\beta$, denoted by $Wei(\alpha_{i}, \beta)$.
\end{enumerate*} The computations  are done for the cases when the strength of the system $K(t)$ is decreasing with time both exponentially and linearly. As before, the values of $c_{T}$, $c_{N}$ and $c_{Z}$ are kept unchanged, i.e. $c_{T} = c_{N} = c_{Z} = 1$, and different choices for the costs incurred from replacement at failure have been considered. 

Under similar distributional assumptions, we have calculated the optimum values $\hat{T}$, $\hat{N}$ and $\hat{Z}$ corresponding to the minimum value of the expected cost rate $C(T, N, Z)$. The results are presented in Table \ref{tab4}.
\begin{center}
	\begin{table}[h!]
		\centering
		\caption{Optimal $\hat{T}$, $\hat{N}$, $\hat{Z}$ and the corresponding minimal expected cost rates $C_{1}(\hat{T})$, $C_{2}(\hat{N})$ and $C_{3}(\hat{Z})$ for independent but not identically distributed damages using the Approx: GS method. Means of the relevant distributions are given in parentheses.}
		\label{tab3}
		\resizebox{\columnwidth}{2.5 cm}{
			\begin{tabular}{c | c | c | c | c c | c c | c c}
				\hline
				$K(t)$ & $F$ & $G_{i}$ & $c_{K}$ & $\hat{T}$ & $C_{1}(\hat{T})$ & $\hat{N}$ & $C_{2}(\hat{N})$ & $\hat{Z}$ & $C_{3}(\hat{Z})$\\
				\hline
				$50\exp(-0.05t)$ & $Exp(2)$ & $Ga(0.5+((i-1)\times0.1),5)$ & $4$ & $1.92$ & $0.725$ & $4$ & $0.571$ & $32.66$ & $0.442$\\
				& $(0.5)$ & $(2.5+(i-1)\times0.1)$ & & & & & & &\\					
				\hline
				$\max\left\lbrace60 - t, 0\right\rbrace$ & $Exp(0.4)$ & $Ga(0.5\times(0.6)^{i-1},5)$ & $4$ & $5.26$ & $0.359$ & $2$ & $0.207$ & $17.37$ & $0.213$\\
				& $(2.5)$ & $(2.5\times(0.6)^{i-1})$ & & & & & & &\\
				\hline
				\hline
				$50\exp(-0.05t)$ & $LN(2, 1)$ & $Wei(10\times(0.6)^{i-1}, 15)$ & $2$ & $17.96$ & $0.059$ & $2$ & $0.065$ & $11.7$ & $0.065$\\
				& $(12.18)$ & $(150\times(0.6)^{i-1})$ & & & & & & &\\
				\hline					
				$\max\left\lbrace60 - t, 0\right\rbrace$ & $LN(2, 1)$ & $Wei(10\times(1.5)^{i-1}, 5)$ & $4$ & $15.13$ & $0.083$ & $2$ & $0.069$ & $13.77$ & $0.069$\\
				& $(12.18)$ & $(50\times(1.5)^{i-1})$ & & & & & & &\\
				\hline
		\end{tabular}}
	\end{table}
\end{center}

\begin{center}
	\begin{table}[]
		\centering
		\caption{Optimal $\hat{T}$, $\hat{N}$, $\hat{Z}$ and the corresponding minimal expected cost rate $C(\hat{T}, \hat{N}, \hat{Z})$ for independent but not identically distributed damages. Means of the relevant distributions are given in parentheses. }
		\label{tab4}
		\resizebox{\columnwidth}{2.5 cm}{
			\begin{tabular}{c | c | c | c | c | c c c c}
				\hline
				$K(t)$ & $F$ & $G_{i}$ & $c_{K}$ & Method & $\hat{T}$ & $\hat{N}$ & $\hat{Z}$ & $C(\hat{T}, \hat{N}, \hat{Z})$\\
				\hline
				$50\exp(-0.05t)$ & $Exp(2)$ & $Ga(0.5+(i-1)\times0.1, 5)$ & $4$ & Approx: GS & $4.91$ & $7$ & $33.97$ & $0.412$\\
				& $(0.5)$ & $(2.5+(i-1)\times0.1)$ &  & Approx: SA &  $4.35$ & $7$ & $35.45$ & $0.425$\\
				\hline
				$\max\left\lbrace 60 - t, 0 \right\rbrace$ & $Exp(0.4)$ & $Ga(0.5\times(0.6)^{i-1},5)$ & $4$ & Approx: GS & $24.02$ & $3$ & $14.83$ & $0.178$\\
				& $(2.5)$ & $(2.5\times(0.6)^{i-1})$ & & Approx: SA & $23.92$ & $3$ & $14.57$ & $0.179$\\
				\hline
				\hline
				$50\exp(-0.05t)$ & $LN(2, 1)$ & $Wei(10\times(0.6)^{i-1}, 15)$ & $2$ & Approx: GS & $21.98$ & $4$ & $16.75$ & $0.054$\\
				& $(12.18)$ & $(150\times(0.6)^{i-1})$ & & Approx: SA &  $22.08$ & $4$ & $16.57$ & $0.053$\\
				\hline
				$\max\left\lbrace 60 - t, 0 \right\rbrace$ & $LN(2, 1)$ & $Wei(10\times(1.5)^{i-1}, 5)$ & $4$ & Approx: GS & $35.20$ & $3$ & $13.97$ & $0.049$\\
				& $(12.18)$ & $(50\times(1.5)^{i-1})$ & & Approx: SA &  $34.35$ & $3$ & $14.08$ & $0.053$\\
				\hline
		\end{tabular}}
	\end{table}
\end{center}

\subsection{Dependent Damage Distribution} 
In order to model dependent damages, a multivariate damage  distribution needs to be considered. We consider a model in which the damage $W_{i}$ due to the $i$th shock can be expressed as $W_{i} = Z_{0} + Z_{i}$, where $Z_0$ is a random variable representing the minimum damage that arrival of a shock can cause to the unit and $Z_i$ is the additional damage caused by the $i$th shock depending on its severity, etc.. Then the successive damages $W_{1}, W_{2},\ldots$ become dependent because of the common minimum damage $Z_{0}$. If the minimum damage $Z_{0}$ and the additional damages $Z_{i}$'s are assumed to be independent $Ga(\theta_{i}, 1)$ random variables for $i = 0, 1, 2,\ldots$, then the joint distribution of $W_{1},\ldots,W_{n}$, for given $n$, is known as the Cheriyan and Ramabhadran's multivariate Gamma distribution \citep{Kotz2000}.
The distribution function of the cumulative damage $U = \sum_{i=1}^{n}W_{i}$ under these assumptions do not fall into any known class of distributions. 
As we have frequently mentioned, there are several other difficulties in evaluating the expected cost rates since the expressions are not in a closed form. By using the simulation method, we can overcome these complications while having less computational burden. In this dependent modeling, in particular, the generation of successive damages is simple due to the additive form of the $W_i$'s. The objective, similar to the previous cases, is to find the optimal values of $T$, $N$ and $Z$, which result in minimum expected cost rates. 

In the following illustrations, as before, we consider the shocks to arrive according to a renewal process with inter-arrival time distribution being 
\begin{enumerate*}[label = (\itshape\roman*\upshape)] 
	\item Exponential distribution with mean $1 / \lambda$, denoted by $Exp(\lambda)$, and \item Log-normal distribution with Normal parameters $\mu$ and $\sigma$, denoted by $LN(\mu, \sigma)$.
\end{enumerate*}
The dependent damages are assumed to follow Cheriyan and Ramabhadran's multivariate Gamma distribution with parameters $\theta_{0}$ and $\theta_{j} = \theta$ for all $j = 1, 2,\ldots$, denoted by $MVGa(\theta_{0}, \theta)$, with mean damage equal to $\theta_{0} + \theta$. The strength of the operating unit can be either exponentially or linearly degrading and the assumptions on the costs incurred from preventive replacement of the unit remains same. The expected cost rate $C(T, N, Z)$ is also minimized as a function of $T$, $N$ and $Z$ taken simultaneously. The computational burden in the  simulation method does not increase much because of the dependent damages. The numerical results  for finding $\hat{T}$, $\hat{N}$ and $\hat{Z}$, separately and simultaneously,  are presented in Tables \ref{tab5} and \ref{tab6}, respectively.

\begin{center}
	\begin{table}[h!]
		\centering
		\caption{Optimal $\hat{T}$, $\hat{N}$, $\hat{Z}$ and the corresponding minimal expected cost rates $C_{1}(\hat{T})$, $C_{2}(\hat{N})$ and $C_{3}(\hat{Z})$ for dependent damage distributions using the Approx: GS method. Means of the relevant distributions are given in parentheses. }
		\label{tab5}
		\resizebox{\columnwidth}{2.5 cm}{
			\begin{tabular}{c | c | c | c | c c | c c | c c}
				\hline
				$K(t)$ & $F$ & $G$ & $c_{K}$ & $\hat{T}$ & $C_{1}(\hat{T})$ & $\hat{N}$ & $C_{2}(\hat{N})$ & $\hat{Z}$ & $C_{3}(\hat{Z})$\\
				\hline
				$100\exp(-0.1t)$ & $Exp(0.2)$ & $MVGa(0.5,10)$ & $2$ & $10.79$ & $0.118$ & $2$ & $0.123$ & $18.91$ & $0.122$\\
				& $(5)$ & $(10.5)$ & & & & & & &\\					
				\hline
				$\max\left\lbrace60 - t, 0\right\rbrace$ & $Exp(0.2)$ & $MVGa(10,5)$ & $4$ & $9.59$ & $0.157$ & $2$ & $0.112$ & $24.31$ & $0.104$\\
				& $(5)$ & $(15)$ & & & & & & &\\
				\hline
				\hline
				$100\exp(-0.03t)$ & $LN(2, 1)$ & $MVGa(0.5,10)$ & $2$ & $28.98$ & $0.042$ & $3$ & $0.041$ & $24.76$ & $0.041$\\
				& $(12.18)$ & $(10.5)$ & & & & & & &\\
				\hline					
				$\max\left\lbrace50 - t, 0\right\rbrace$ & $LN(2, 1)$ & $MVGa(0.5,5)$ & $2$ & $27.61$ & $0.043$ & $3$ & $0.049$ & $12.91$ & $0.05$\\
				& $(12.18)$ & $(5.5)$ & & & & & & &\\
				\hline
		\end{tabular}}
	\end{table}
\end{center}

\begin{center}
	\begin{table}[]
		\centering
		\caption{Optimal $\hat{T}$, $\hat{N}$, $\hat{Z}$ and the corresponding minimal expected cost rate $C(\hat{T}, \hat{N}, \hat{Z})$ for dependent damages. Means of the relevant distributions are given in parentheses. }
		\label{tab6}
		\resizebox{\columnwidth}{2.5 cm}{
			\begin{tabular}{c | c | c | c | c | c c c c}
				\hline
				$K(t)$ & $F$ & $G$ & $c_{K}$ & Method & $\hat{T}$ & $\hat{N}$ & $\hat{Z}$ & $C(\hat{T}, \hat{N}, \hat{Z})$\\
				\hline
				$100\exp(-0.1t)$ & $Exp(0.2)$ & $MVGa(0.5,10)$ & $2$ & Approx: GS & $13.62$ & $4$ & $24.88$ & $0.099$\\
				& $(5)$ & $(10.5)$ &  & Approx: SA &  $13.59$ & $4$ & $25.57$ & $0.099$\\
				\hline
				$\max\left\lbrace 60 - t, 0 \right\rbrace$ & $Exp(0.2)$ & $MVGa(10,5)$ & $4$ & Approx: GS & $25.40$ & $3$ & $22.49$ & $0.088$\\
				& $(5)$ & $(15)$ & & Approx: SA & $25.10$ & $3$ & $23.11$ & $0.088$\\
				\hline
				\hline
				$100\exp(-0.03t)$ & $LN(2, 1)$ & $MVGa(0.5,10)$ & $2$ & Approx: GS & $40.19$ & $4$ & $29.71$ & $0.035$\\
				& $(12.18)$ & $(10.5)$ & & Approx: SA &  $41.61$ & $4$ & $28.46$ & $0.035$\\
				\hline
				$\max\left\lbrace 50 - t, 0 \right\rbrace$ & $LN(2, 1)$ & $MVGa(0.5,5)$ & $2$ & Approx: GS & $33.89$ & $4$ & $16.10$ & $0.037$\\
				& $(12.18)$ & $(5.5)$ & & Approx: SA &  $33.94$ & $4$ & $16.01$ & $0.039$\\
				\hline
		\end{tabular}}
	\end{table}
\end{center}

\section{Case Studies}\label{case}
\begin{figure}
 \centering
 \includegraphics[scale=0.45]{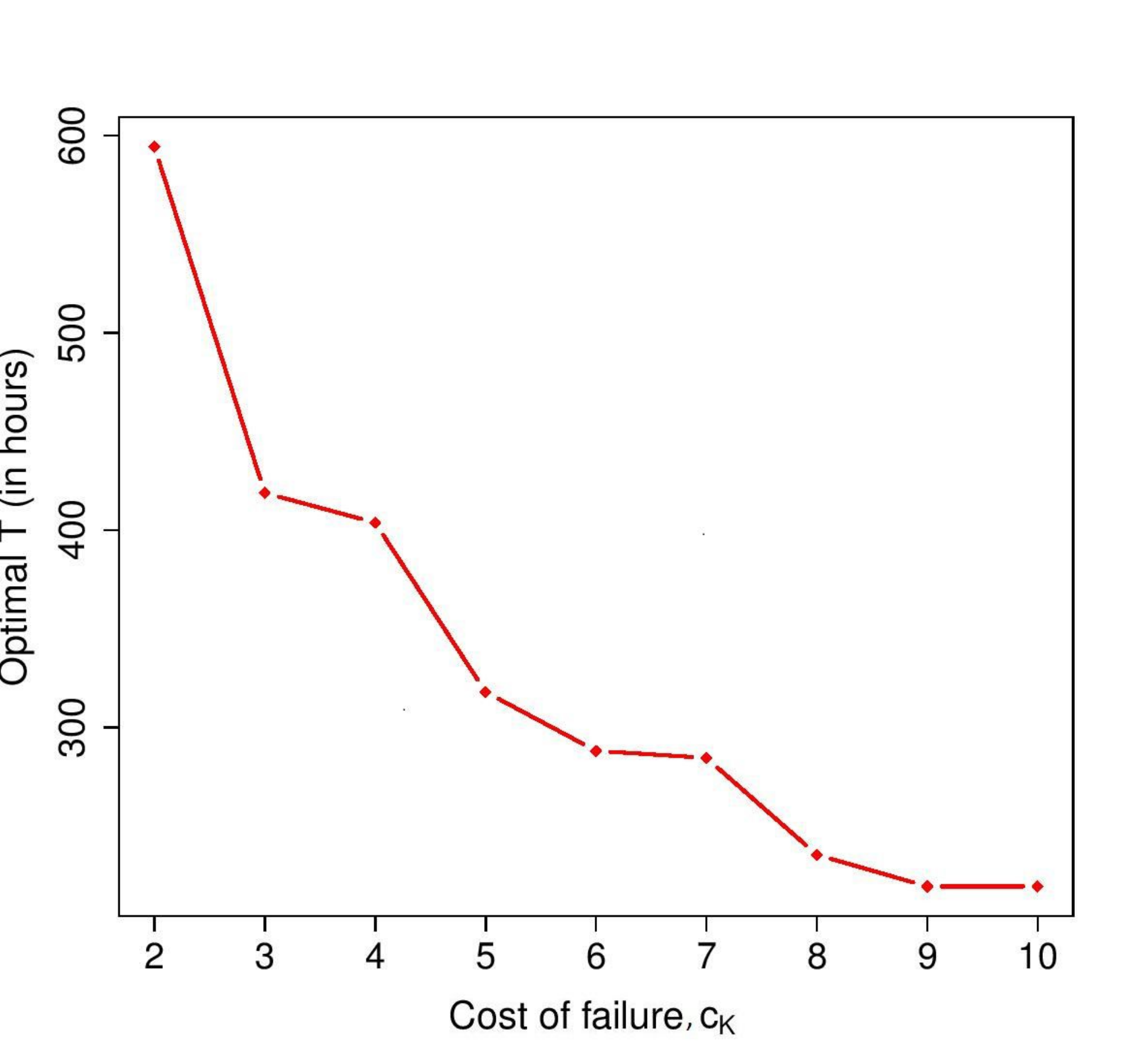}
\caption{The optimal $T$ for mailbox experiment keeping cost of replacement fixed at unity.}
\label{T_graph}
\end{figure}

\begin{figure}
 \centering
 \includegraphics[scale=0.45]{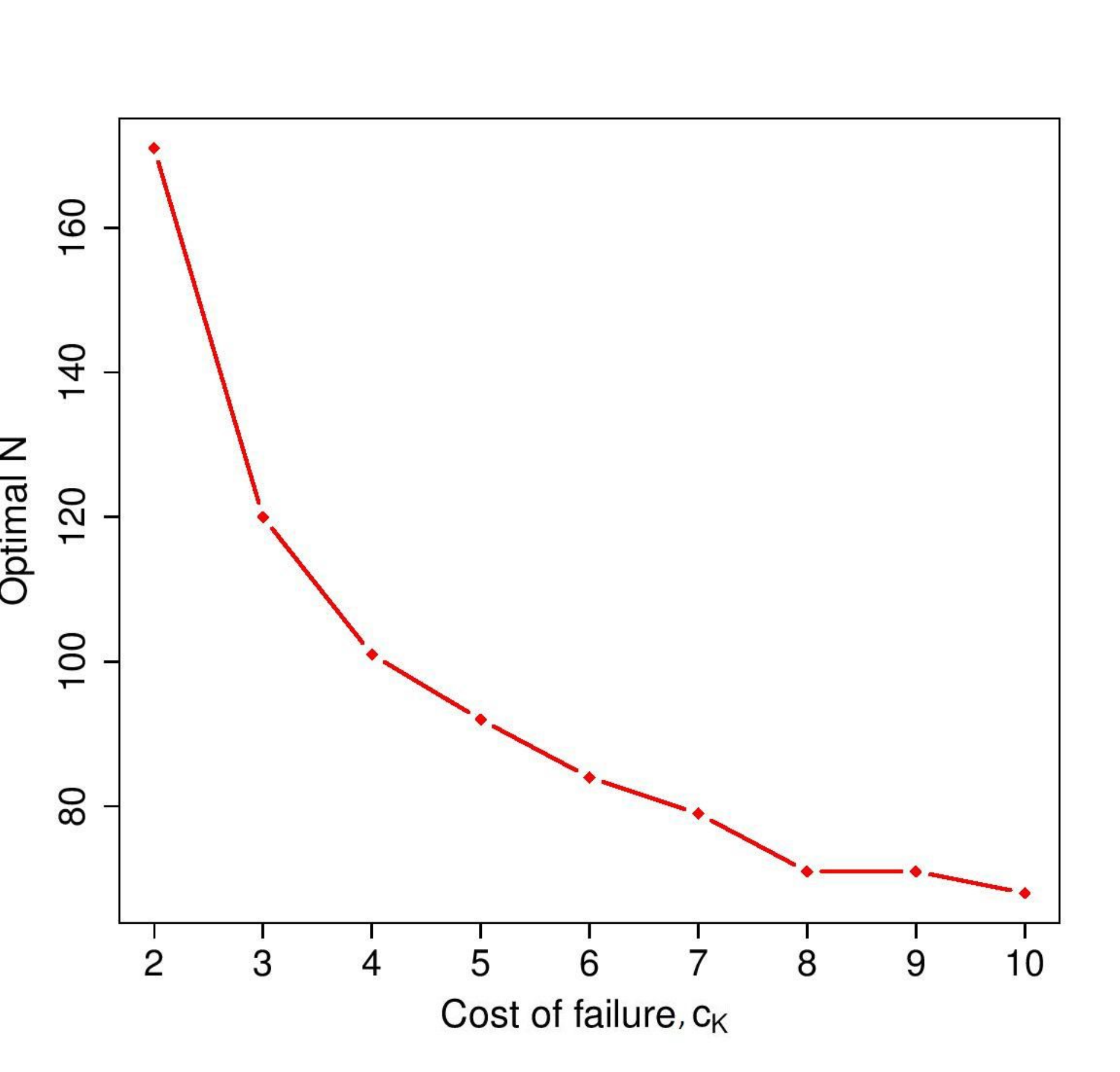}
\caption{The optimal $N$ for mailbox experiment keeping cost of replacement fixed at unity.}
\label{N_graph}
\end{figure}

\begin{figure}
 \centering
 \includegraphics[scale=0.45]{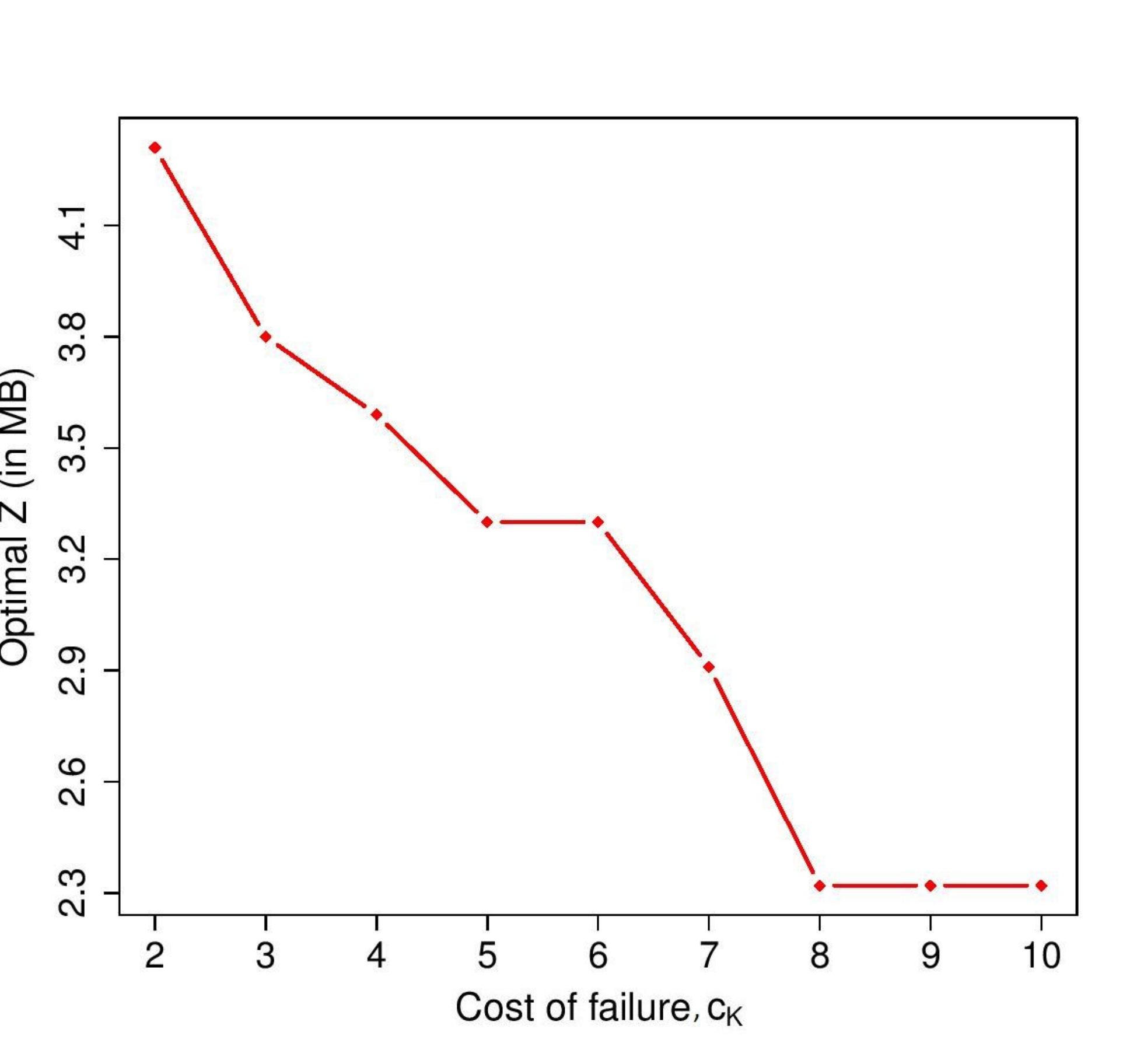}
\caption{The optimal $Z$ for mailbox experiment keeping cost of replacement fixed at unity.}
\label{Z_graph}
\end{figure}

\begin{figure}
 \centering
 \includegraphics[scale=0.45]{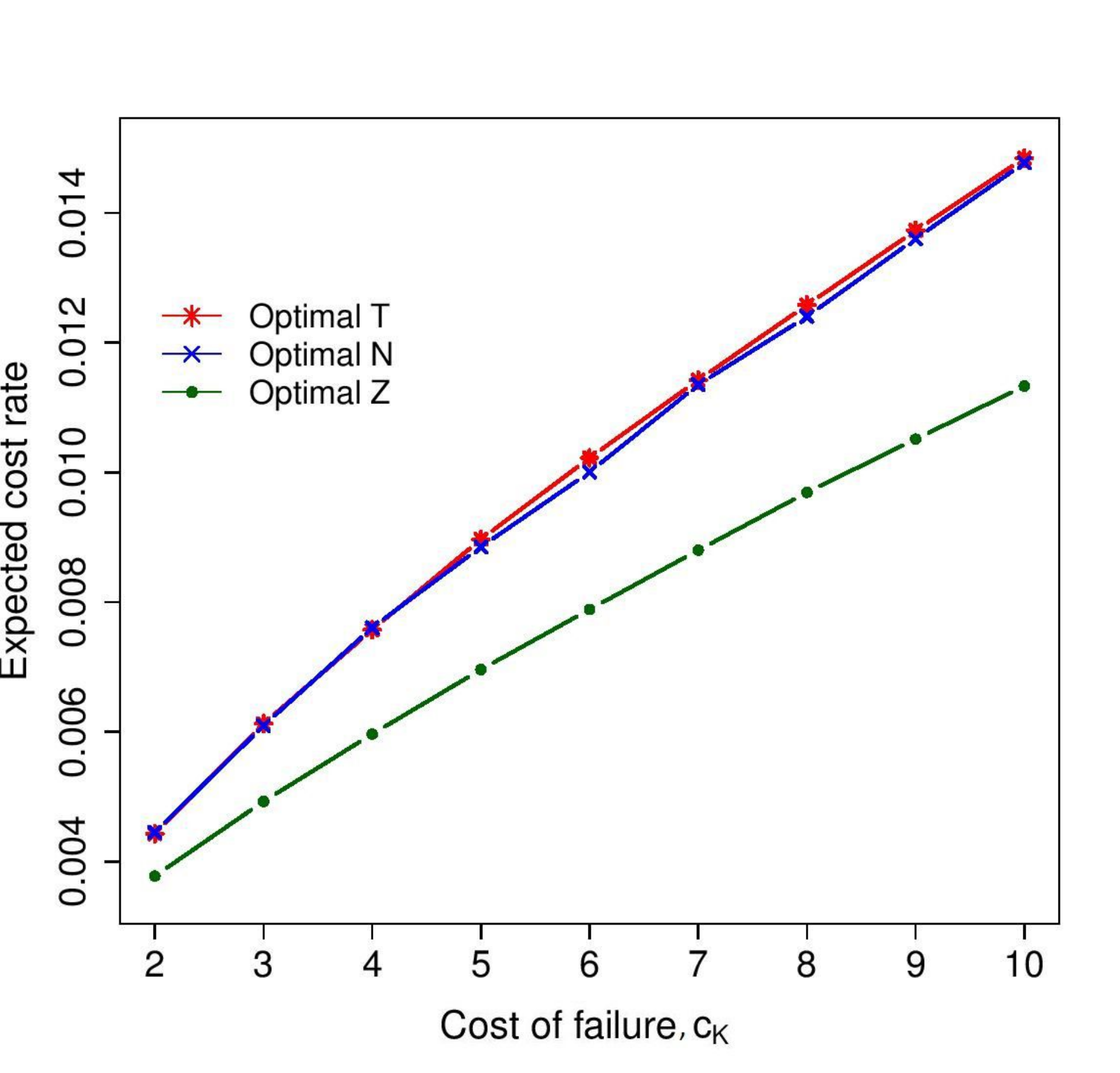}
\caption{Comparison of expected cost rates corresponding to optimal $T$, $N$, $Z$}
\label{Mail}
\end{figure}

In Section \ref{Intro}, we have discussed some application in database management systems for its efficient operation. It is a common practice among email users to forward emails automatically from various email accounts to a preferred email account for ease of operation. In this process, users normally do not clean the mailbox of the secondary email accounts. As a result of accumulation of emails over time, the secondary mailbox becomes full and the account fails to receive any further email. \citet{Bhuyan2017b} collected data on $22$ such identical systems and observed failure time (in hours) data and the number of emails received up to the time of failure. The mailbox limit (that is, the strength of the system $s(t)$) is kept fixed at $5$ MB. In a preliminary data analysis, the average number of arriving shocks seems to increase with time in a linear fashion. Therefore, we assume that emails arrive according to a homogeneous Poisson process and the estimated mean inter-arrival time is $3$ hours $27$ minutes. We find that the Log-normal distribution fits the size (in MB) of the successive emails well and the corresponding parameter estimates are $\hat{\mu}=-7.32$ and $\hat{\sigma}=3.16$ with mean 97.57 KB. The optimal values of $T$, $N$, and $Z$ are obtained by minimizing $\hat{C_{1}}(T)$, $\hat{C_{2}}(N)$, and $\hat{C_{3}}(Z)$, respectively, and plotted against $c_K$ in Figures \ref{T_graph}-\ref{Z_graph}, keeping the other cost components $c_T, c_N$ and $c_Z$ for replacement fixed at unity. As expected, the optimal value of $T$ decreases as $c_{K}$ increases and sharp decline is observed up to $c_{K}=5$ . Similar patterns are also observed for optimal values of $N$ and $Z$. The expected cost rates corresponding to optimal replacement strategies are plotted against $c_{K}$ in Figure \ref{Mail}. It is observed that the optimal strategy based on $Z$ is better compared to the same based on $T$ and $N$ with respect to the expected cost rate. As for illustration by considering all the design parameters together, the optimal replacement strategy $(\hat{T},\hat{N},\hat{Z})=(708.89,183,3.86)$ and the associated expected cost rate $3.82\times10^{-3}$ are obtained by minimizing $\hat{C}(T,N,Z)$ for $c_K=2$ with $c_T=c_N=c_Z=1$. 

\begin{figure}
 \centering
 \includegraphics[scale=0.45]{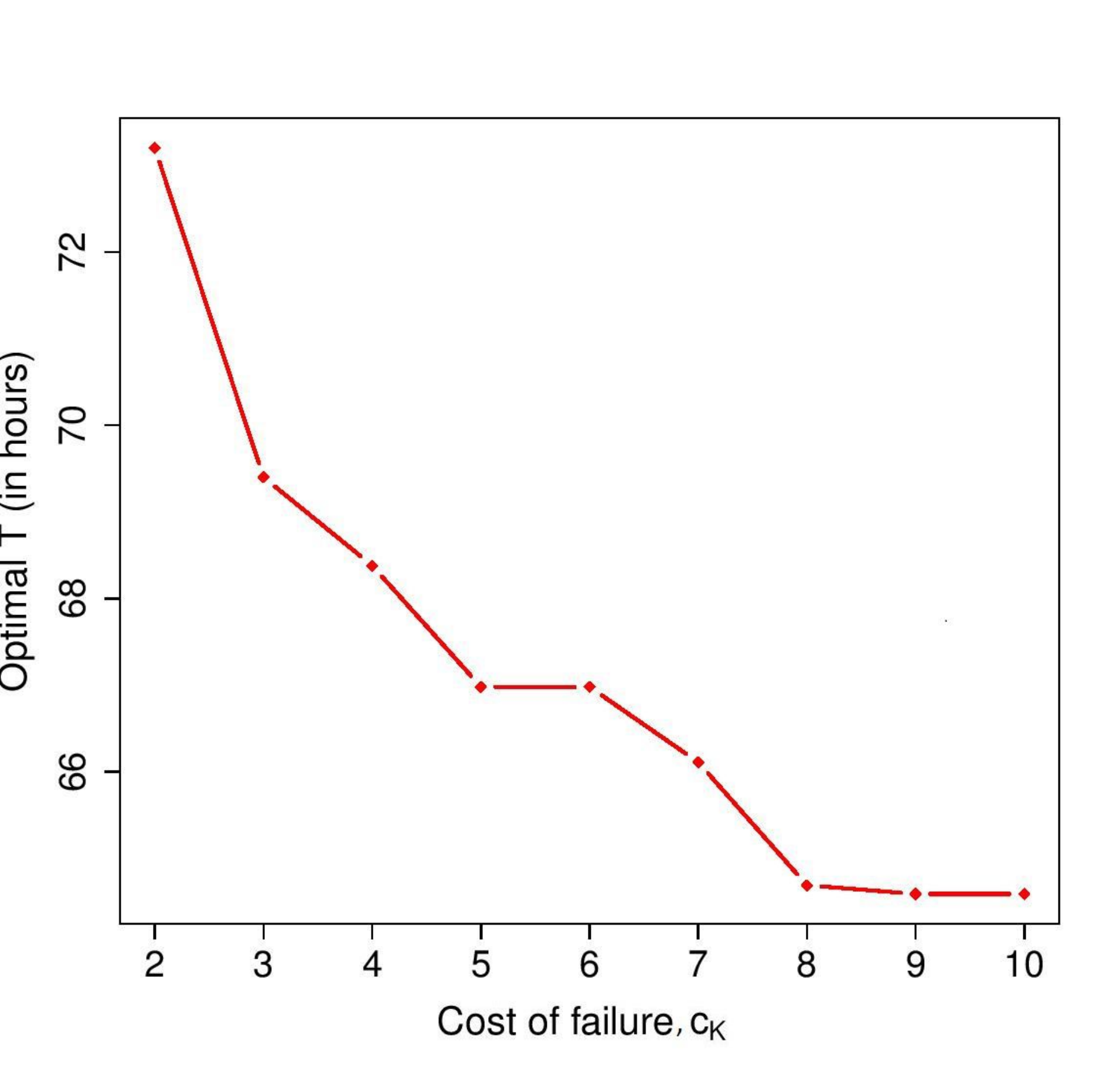}
\caption{The optimal $T$ for cell-phone battery experiment keeping cost of replacement fixed at unity.}
\label{TT_graph}
\end{figure}

\begin{figure}
 \centering
 \includegraphics[scale=0.45]{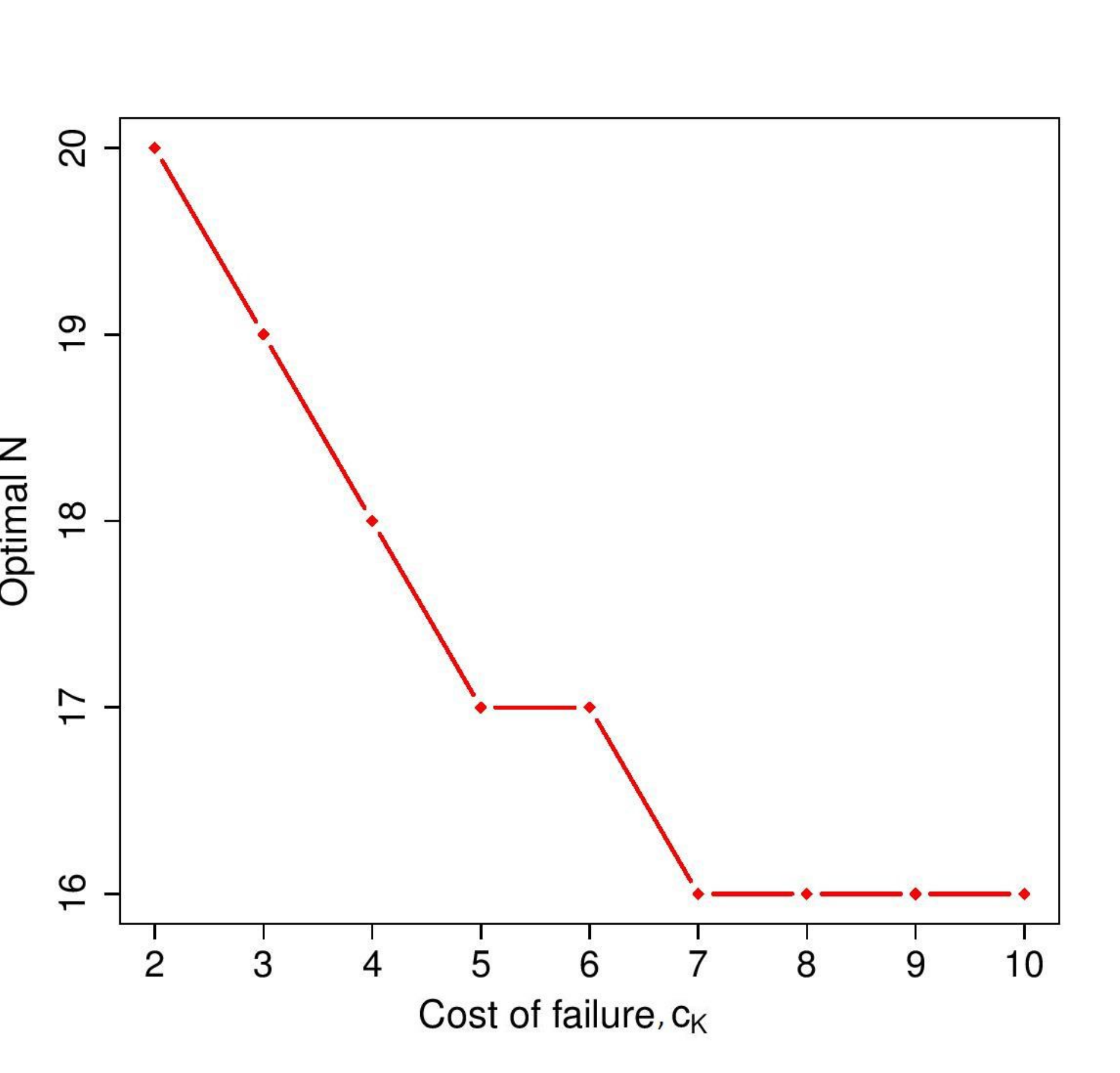}
\caption{The optimal $N$ for cell-phone battery experiment keeping cost of replacement fixed at unity.}
\label{NN_graph}
\end{figure}

\begin{figure}
 \centering
 \includegraphics[scale=0.45]{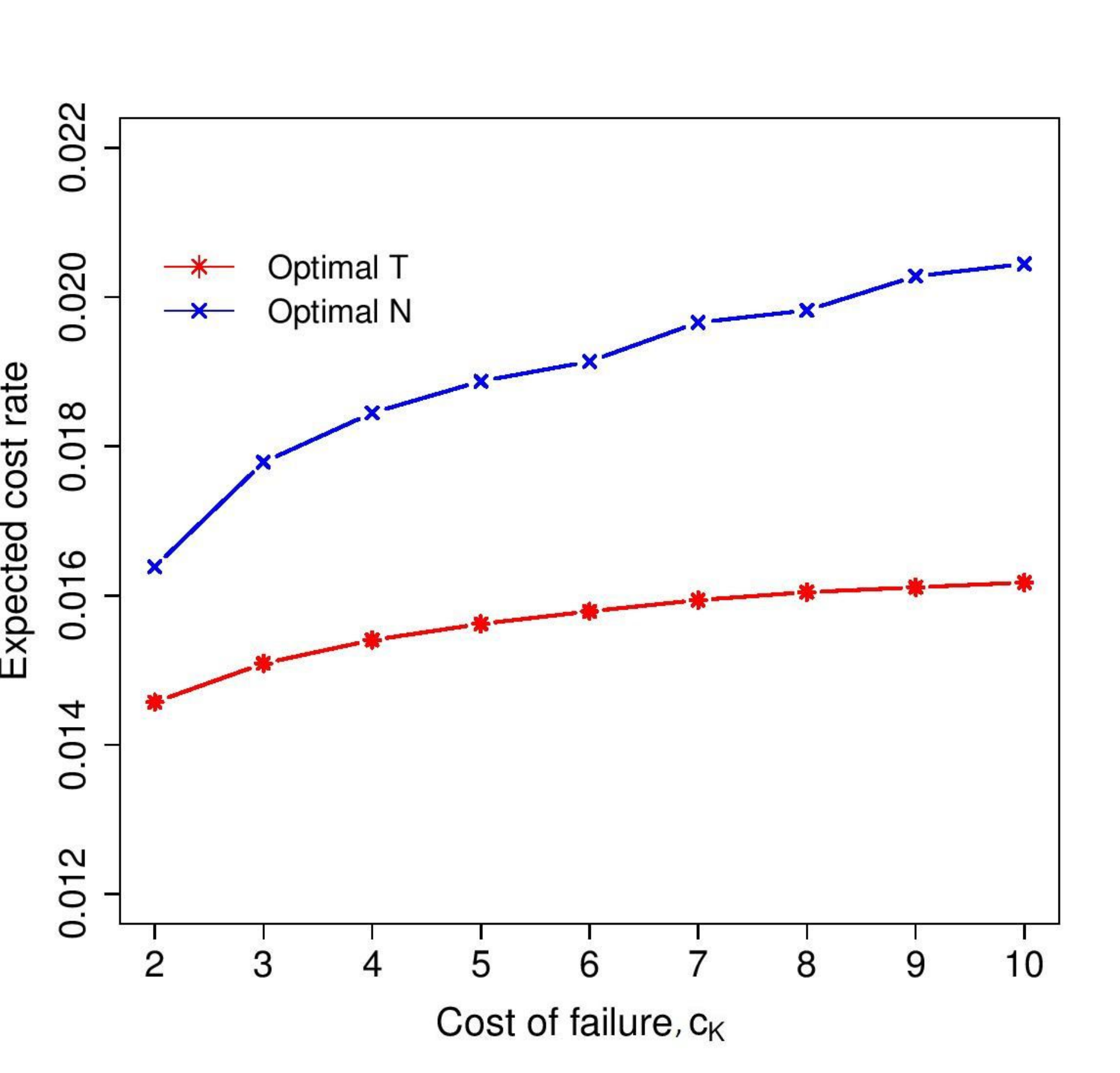}
\caption{Comparison of expected cost rates corresponding to optimal $T$ and $N$}
\label{Cell}
\end{figure}

As discussed before, the electric power of a dry cell or battery, initially stored by chemical energy, is weakened by continuous oxidation process and is subject to frequent use leading to accumulated damage or energy loss. Similar phenomenon happens in cell-phone battery after each and every recharge. Once fully charged, cell-phone battery looses its energy over time due to normal functionalities of the cell-phone in the switch-on mode and frequent incoming and outgoing calls leading to accumulated damage or energy loss. \citet{Bhuyan2017b} analysed data on $11$ identical cell-phone batteries based on failure time (in hours) data and the number of calls (incoming and outgoing) up to the time of failure. We assume that incoming and outgoing calls take place according to a homogeneous Poisson process with the estimated rate $\hat{\lambda}=0.29$. We also assume that $K(t)=A\exp(-Bt)$ and the damages due to successive calls follow iid Gamma distribution. 
For identifiability, the initial strength $A$ is fixed at $100$. See \citet{Bhuyan2017b} for more details. Nevertheless, we consider the estimated scale parameter $\hat{\theta}=1.54$ and shape parameter $\hat{\delta}=0.193$ of the Gamma damages and $\hat{B}=0.041$ to carry out our analysis. The optimal values $\hat{T}$ and $\hat{N}$ are obtained by minimizing $\hat{C_{1}}(T)$ and $\hat{C_{2}}(N)$, respectively, and plotted against $c_K$ in Figures \ref{TT_graph}-\ref{NN_graph}, keeping the cost of replacement fixed at unity. As expected, $\hat{T}$ decreases as $c_{K}$ increases. A sharp decline is observed for $\hat{N}$ up to $c_{K}=5$. Note that the accumulated energy consumption due to incoming and outgoing calls are not observable. Therefore, we do not provide any optimal replacement strategy based on $Z$. The expected cost rates corresponding to the replacement strategies based on $T$ and $N$ are plotted against $c_{K}$ in Figure \ref{Cell}. It is observed that the optimal strategy based on $T$ is much better compared to that based on $N$. Now, considering these two design parameters together, the optimal replacement strategy 
$(\hat{T},\hat{N})=(73.41 \text{hours}, 28)$ 
and the associated expected cost rate $1.458\times10^{-2}$, 
are obtained by minimizing $\hat{C}(T,N,\infty)$ for $c_K=2$ with $c_T=c_N=1$.

\section{Concluding Remarks}\label{sec7}
The cumulative damage model with strength degradation unlike that with a fixed strength is more common and realistic. However, the replacement problem under such model with decreasing strength has not yet been addressed. The unit is preventively replaced before failure at a scheduled time $T$, shock number $N$ and a damage level $Z$ whichever occurs first and correctively replaced at failure. Under this replacement policy, we have obtained the expressions for the expected cost rates of replacement at $T$, $N$ and $Z$ individually, or all taken together. These expressions are not in closed form which makes it extremely difficult to analytically derive the optimum policy. Besides, evaluating the convolutions of the distribution functions itself is a complicated process. 
In this work, probably for the first time, the computational issues associated with the replacement problem for cumulative damage model with degrading strength has been discussed. We have proposed a simulation algorithm for evaluating the expected cost rates. The method of simulation reduces the computational burden while providing room for a wider range of distributional choices. We have also considered some generalized cases where the damages caused by shocks can either be dependent or independent but not identically distributed. In fact, even for a general (that is, non-renewal) point process modeling for the shock arrivals, the simulation method can be readily implemented as long as the shock arrival process can be simulated. 

In many real life scenarios, shocks appear from multiple sources thereby causing damages with different distributions depending on the source of the corresponding shock. One can then ideally model the damage distribution corresponding to a shock to have a mixture distribution. This mixture damage distribution does not generally have closed form expression for convolutions and, therefore, finding optimal replacement strategy is computationally difficult. The proposed algorithm can handle such cases easily. Furthermore, the proposed algorithm is readily generalised to find optimal strategy based on $Z$ and/or $N$ as a function of time, which may be useful in real implementations under dynamic stress-strength interference.

Again, in many real situations, initial strength or its path of deterioration over time is random. Sometimes, deterioration of strength over time is due to various environmental causes changing stochastically at every instant. Another possible scenario is that the strength of the operating unit degrades in a non-monotonic fashion. The unit may go through some auto-repairing process that will cause some ups and downs in its strength \citep{Ebrahimi1993}. Evaluation of the expected cost rates for replacement in those cases are complicated which adds to the reasons why simulation method should be preferred over other competing methods.

Note that, in view of obtaining the optimal $T$, $N$ and $Z$ for preventive replacement, one can also determine, at least in theory, the expected time for corrective replacement at failure in the presence/absence of any of the four preventive replacement plans. This may be of interest to the industry concerned in view of the huge cost involved with a system failure. One can design a simulation algorithm for this purpose in line of that described in Section \ref{sec4}. The assumption of immediate replacement used in the present work  may not be a reasonable one in practice. Incorporation of delayed replacement will require not only the modeling of the distribution of duration of replacement, but also the information on the cost of down time for replacement due to loss of profit, service, etc., during this period of random length. This may be theoretically challenging even when all this information is available. A simulation based method may be the solution in such a complicated scenario.

 \renewcommand{\theequation}{A\arabic{equation}}    
\setcounter{equation}{0}  
\section*{Appendix}

\subsection*{Derivation for replacement at time $T$}
The probability of preventive replacement $p_{T}$ due to reaching age $T$ prior to failure occurrence  can be obtained as
\begin{align}\label{eq3.1}
\begin{split}
p_{T} & = \sum_{j=0}^{\infty} P\left[ S_{j} \leq T, S_{j+1} > T, W_{0} + W_{1} + \cdots + W_{j} < K(T)\right]\nonumber\\
& = P\left[X_{1} > T\right] + \sum_{j=1}^{\infty} P\left[ S_{j} \leq T, S_{j+1} > T, W_{1} + \cdots + W_{j} < K(T)\right]\\
& = \bar{F}(T) + \sum_{j=1}^{\infty} \left[ F^{(j)}(T) - F^{(j+1)}(T) \right]G^{(j)}(K(T)),\nonumber
\end{split}
\end{align}
where $S_{0}~=~W_{0}~=~0$. Since the unit is replaced either at the planned time $T$ or at failure, the probability that the unit is replaced at failure is given by $p_{K} = 1 - p_{T}$. If $c_{T}$ and $c_{K}$ are the costs incurred when the unit is replaced at $T$ and at failure, respectively, then the expected cost of replacement can be written as
\begin{equation}\label{eq3.2}
\tilde{C}_{1}(T) = c_{K} - \left(c_{K} - c_{T}\right) \left[\bar{F}(t) + \sum_{j=1}^{\infty} \left[ F^{(j)}(T) - F^{(j+1)}(T) \right]G^{(j)}(K(T)) \right].
\end{equation}
If $S$ denotes the time to replacement, then for any $t \in (0, T)$,
\begin{align*}
P\left[ S > t \right] & = \sum_{j=0}^{\infty} P\left[ S_{j} \leq t , S_{j+1} > t, W_{0} + W_{1} + \cdots + W_{j} < K(t) \right]\\
& = \bar{F}(t) + \sum_{j=1}^{\infty} \left[F^{(j)}(t) - F^{(j+1)}(t)\right]G^{(j)}(K(t)).
\end{align*}
Further, $P[S>t]=0$ for $t\ge T$. Then, the mean time to replacement for this case is given by
\begin{equation}\label{eq3.3}
\int_{0}^{\infty} P\left[ S > t \right]dt = \int_{0}^{T}\bar{F}(t)dt + \sum_{j=1}^{\infty} \int_{0}^{T} \left[F^{(j)}(t) - F^{(j+1)}(t)\right]G^{(j)}(K(t)) dt.
\end{equation}
Then, dividing the expected cost for replacement (\ref{eq3.2}) by the mean time to replacement (\ref{eq3.3}), we get the expected cost rate given by (\ref{eq3.4}). 

\subsection*{Derivation for replacement at the shock number $N$}
The probability that the unit is replaced at the $N$th shock prior to failure occurrence is
\begin{align}\label{eq3.5}
\begin{split}
p_{N} & = P\left[ \sum_{j=0}^{N} W_{j} < K(S_{N}) \right]\nonumber\\
& = \int_{0}^{\infty} G^{(N)}(K(s)) dF^{(N)}(s).
\end{split}
\end{align}
Similar to the previous case, the probability of replacement at failure is given by $p_{K} = 1 - p_{N}$. The costs of replacement at the $N$th shock and at failure are assumed to  be $c_{N}$ and $c_{K}$, respectively. Then the expected cost $\tilde{C}_{2}(N)$ of replacement can be written as
\begin{equation}\label{eq3.6}
\tilde{C}_{2}(N) = c_{K} - \left(c_{K} - c_{N}\right)\int_{0}^{\infty} G^{(N)}(K(s)) dF^{(N)}(s).
\end{equation}
For any $t \in [0, \infty)$, the  probability that the unit is not replaced before time $t$ is given by,
\begin{align*}
\begin{split}
P\left[ S > t \right] & = \sum_{j=0}^{N-1} P\left[ S_{j} \leq t, S_{j+1} > t, W_{0} + W_{1} + \cdots + W_{j} < K(t) \right] \\
& = P\left[X_{1} > t\right] + \sum_{j=1}^{N-1} P\left[ S_{j} \leq t, S_{j+1} > t, W_{1} + \cdots + W_{j} < K(t) \right]\\
& = \bar{F}(t) + \sum_{j=1}^{N-1} \left[F^{(j)}(t) - F^{(j+1)}(t)\right]G^{(j)}(K(t))
\end{split}
\end{align*}
Then the mean time to replacement in this case will be
\begin{equation}\label{eq3.7}
\int_{0}^{\infty} P\left[ S > t \right] dt = \mu_{F} + \sum_{j=1}^{N-1} \int_{0}^{\infty} \left[F^{(j)}(t) - F^{(j+1)}(t)\right]G^{(j)}(K(t)) dt, 
\end{equation}
where $\mu_{F} = E\left[X_{i}\right],~i = 1,2,\ldots$.
Then, dividing the expected cost for replacement (\ref{eq3.6}) by the mean time to replacement (\ref{eq3.7}), we get the expected cost rate (\ref{eq3.8}).

\subsection*{Derivation for replacement at the cumulative damage $Z$}
The problem of replacement at $Z$ needs to be looked at in a bit different way from those for replacement at time $T$ or at shock number $N$. Let $T_{0}$ be the time such that $K(T_{0}) = Z$. Thus, before $T_{0}$, the replacement of the unit can either be due to the damage level $Z$ or due to failure; but after $T_{0}$ the replacement will be only due to failure of the unit. As discussed before, we assume that the replacement is corrective rather than preventive, if the accumulated damage exceeds both $Z$ and the strength at the time of a shock arrival. The probability $p_{Z}$ that the replacement is done due to damage $Z$ prior to failure occurrence is, therefore, obtained as
\begin{eqnarray}\label{eq3.9}
p_{Z} &=& \sum_{j=0}^{\infty} P\left[ W_{0} + W_{1} + \cdots + W_{j} < Z \leq W_{0} + W_{1} + \cdots + W_{j+1} < K(S_{j+1}),\ S_{j+1} < T_0 \right]\nonumber\\
&=& \sum_{j=0}^{\infty} \int_{0}^{T_{0}} P\left[ W_{0} + W_{1} + \cdots + W_{j} <  Z \leq W_{0} + W_{1} + \cdots + W_{j+1} < K(s) \right] dF^{(j+1)}(s)\nonumber\\
&=& \int_{0}^{T_{0}} P\left[Z \leq W_{1} < K(s)\right] dF(s) \nonumber\\ 
&&+ \sum_{j=1}^{\infty} \int_{0}^{T_{0}} \int_{0}^{Z} P\left[ Z - x \leq W_{j+1} < K(s) - x \right] dG^{(j)}(x) dF^{(j+1)}(s)\nonumber\\
&=& \int_{0}^{T_{0}} \left[G(K(s)) - G(Z)\right] dF(s) + \sum_{j=1}^{\infty} \int_{0}^{T_{0}} \int_{0}^{Z} \left[ G(K(s) - x) - G(Z - x) \right] dG^{(j)}(x) dF^{(j+1)}(s) \nonumber.
\end{eqnarray}
The replacement is done either at damage level $Z$ or at failure. Therefore, as before, the expected cost of replacement can be written as
\begin{eqnarray}\label{eq3.10}
\tilde{C}_{3}(Z) & = & c_{K} - \left(c_{K} - c_{Z}\right)\Big[ \int_{0}^{T_{0}} \left[G(K(s)) - G(Z)\right] dF(s) \nonumber \\ && + \sum_{j=1}^{\infty} \int_{0}^{T_{0}} \int_{0}^{Z} \left[ G(K(s) - x) - G(Z - x) \right] dG^{(j)}(x) dF^{(j+1)}(s) \Big]
\end{eqnarray}
where $c_{K}$ and $c_{Z}$ are the costs incurred from replacement at $Z$ and at failure, respectively. In order to calculate the mean time to replacement, we proceed by first calculating the probability that the unit is not replaced before some time $t$. To serve our purpose, we need to define a modified time-dependent replacement level $\tilde{K}(t)$ as given by 
\begin{equation}\label{eq3.11}
\tilde{K}(t) = \begin{cases}
Z, & t \leq T_{0}\nonumber\\
K(t), & t > T_{0}.\nonumber
\end{cases}
\end{equation}
Then the probability that replacement is not done during $[0, t]$ will be
\begin{align*}
\begin{split}
P\left[ S > t \right] & = \sum_{j=0}^{\infty} P\left[ S_{j} \leq t, S_{j+1} > t, W_{0} + W_{1} + \cdots + W_{j} < \tilde{K}(t) \right]\\
& = P\left[X_{1} > t\right] + \sum_{j=1}^{\infty} P\left[ S_{j} \leq t, S_{j+1} > t, + W_{1} + \cdots + W_{j} < \tilde{K}(t) \right]\\
& = \bar{F}(t) + \sum_{j=1}^{\infty} \left[F^{(j)}(t) - F^{(j+1)}(t)\right]G^{(j)}(\tilde{K}(t)).
\end{split}
\end{align*}
Therefore, the mean time to replacement is given by
\begin{equation}\label{eq3.12}
\int_{0}^{\infty} P\left[ S > t \right]dt = \mu_{F} + \sum_{j=1}^{\infty} \int_{0}^{\infty} \left[F^{(j)}(t) - F^{(j+1)}(t)\right]G^{(j)}(\tilde{K}(t)) dt.
\end{equation}
Then, dividing the expected cost for replacement (\ref{eq3.9}) by the mean time to replacement (\ref{eq3.12}), we get the expected cost rate (\ref{eq3.13}). 

\subsection*{Derivation for Replacement under simultaneous consideration of $T$, $N$ and $Z$}
It is reasonable to restrict the design space of $(T,N,Z)$ into those choices of $T$ and $Z$ such that $Z\le K(T)$, or $T\le T_0$, so that the  replacement due to $Z$ remains a possibility. As before, let us write $p_{T}$, $p_{N}$, $p_{Z}$ and $p_{K}$ as the probabilities that the unit is replaced at scheduled time $T$, shock number $N$, damage level $Z$ and at failure, respectively. Then
\begin{align}\label{eq3.14}
\begin{split}
p_{T} & = \sum_{j=0}^{N-1} P\left[ S_{j} \leq T, S_{j+1} > T, W_{0} + W_{1} + \cdots + W_{j} < Z\right]\nonumber\\
& = P\left[X_{1} > T\right] + \sum_{j=1}^{N-1} P\left[ S_{j} \leq T, S_{j+1} > T, W_{1} + \cdots + W_{j} < Z\right]\\
& = \bar{F}(T) + \sum_{j=1}^{N-1} \left[ F^{(j)}(T) - F^{(j+1)}(T) \right]G^{(j)}(Z),\nonumber
\end{split}
\end{align}
and
\begin{equation}\label{eq3.15}
p_{N} = F^{(N)}(T) G^{(N)}(Z). \nonumber
\end{equation}
Note that the above expressions of $p_{T}$ and $p_{N}$ are exactly same as those obtained in the case of cumulative damage model with fixed strength \citep[ch-3]{Nakagawa2007}. The probability that the unit is replaced at damage level $Z$ can be calculated as
\begin{eqnarray}\label{eq3.16}
p_{Z} &=& \sum_{j=0}^{N-1} P\left[ W_{0} + W_{1} + \cdots + W_{j} < Z \leq W_{0} + W_{1} + \cdots + W_{j+1} < K(S_{j+1}), S_{j+1} \le T \right]\nonumber\\
&=& \sum_{j=0}^{N-1} \int_{0}^{T} P\left[ W_{0} + W_{1} + \cdots + W_{j} < Z \leq W_{0} + W_{1} + \cdots + W_{j+1} < K(s) \right] dF^{(j+1)}(s)\nonumber\\
&=& \int_{0}^{T} P\left[Z \leq W_{1} < K(s)\right]dF(s) \nonumber \\
&&+ \sum_{j=1}^{N-1} \int_{0}^{T} \int_{0}^{Z} P\left[ Z - x \leq  W_{j+1} < K(s) - x \right] dG^{(j)}(x) dF^{(j+1)}(s)\nonumber\\
&=& \int_{0}^{T} \left[G(K(s)-G(z))\right]dF(s) + \sum_{j=1}^{N-1} \int_{0}^{T} \int_{0}^{Z} \left[ G(K(s) - x) - G(Z - x) \right] dG^{(j)}(x) dF^{(j+1)}(s).\nonumber
\end{eqnarray}
Similarly, the probability that the unit is replaced at failure is
\begin{eqnarray}\label{eq3.17}
p_{K} & =& \sum_{j=0}^{N-1} P\left[ W_{0} + W_{1} + \cdots + W_{j} < Z,  W_{0} + W_{1} + \cdots + W_{j+1} \ge K(S_{j+1}), S_{j+1} \leq T \right]\nonumber\\
& =& \sum_{j=0}^{N-1} \int_{0}^{T} P\left[ W_{0} + W_{1} + \cdots + W_{j} < Z,  W_{0} + W_{1} + \cdots + W_{j+1} \ge K(s)\right] dF^{(j+1)}(s)\nonumber\\
& =& \int_{0}^{T} P[W_{1} \geq K(s)] dF(s) + \sum_{j=1}^{N-1} \int_{0}^{T} \int_{0}^{Z} P\left[ W_{j+1} \ge K(s) - x \right] dG^{(j)}(x) dF^{(j+1)}(s)\nonumber\\
& =& \int_{0}^{T} \bar{G}(K(s))dF(s) + \sum_{j=1}^{N-1} \int_{0}^{T} \int_{0}^{Z} \bar{G}(K(s) - x) dG^{(j)}(x) dF^{(j+1)}(s).\nonumber
\end{eqnarray}
It can be easily verified that $p_{T} + p_{N} + p_{Z} + p_{K} = 1$. Again, write $c_{T}$, $c_{N}$, $c_{Z}$ and $c_{K}$ as the costs of replacement at the planned time $T$, shock number $N$, damage level $Z$ and at failure, respectively, with $c_{K}$ being the largest. Then the expected cost of replacement of the unit is given by
\begin{eqnarray}\label{eq3.18}
\tilde{C}(T, N, Z) & =& c_{K} - \left(c_{K} - c_{T}\right)\Big[\bar{F}(T) + \sum_{j=1}^{N-1} \left\lbrace F^{(j)}(T) - F^{(j+1)}(T) \right\rbrace G^{(j)}(Z)\Big] \nonumber \\ &&- \left(c_{K} - c_{N}\right) F^{(N)}(T) G^{(N)}(Z)
- \left(c_{K} - c_{Z}\right)\Big[\int_{0}^{T} \left[G(K(s)-G(z))\right]dF(s) \nonumber \\ && + \sum_{j=1}^{N-1} \int_{0}^{T} \int_{0}^{Z} \left[ G(K(s) - x) - G(Z - x) \right] dG^{(j)}(x) dF^{(j+1)}(s)\Big].
\end{eqnarray}
For any $t \in [0, T)$, $P\left[ S > t \right]$ is same as the probability that at most $N-1$ shocks occur during $[0, t)$ and the total damage due to those shocks does not  equal or exceed the damage level $Z$. Hence, for any $t \in [0, T)$,
\begin{align*}
P\left[ S > t \right] & = \sum_{j=0}^{N-1} P\left[ S_{j} \leq t, S_{j+1} > t, W_{0} + W_{1} + \cdots + W_{j} < Z \right]\\
& = P\left[X_{1} > t\right] + \sum_{j=1}^{N-1} P\left[ S_{j} \leq t, S_{j+1} > t, W_{1} + \cdots + W_{j} < Z \right]\\
& = \bar{F}(t) + \sum_{j=1}^{N-1} G^{(j)}(Z) \left[ F^{(j)}(t) - F^{(j+1)}(t) \right].
\end{align*}
Since the operating unit is anyway going to be replaced after the planned time $T$, the survival function of $S$ can be written as
\begin{equation*}
P\left[ S > t \right] = \begin{cases}
\bar{F}(t) + \sum_{j=1}^{N-1} G^{(j)}(Z) \left[ F^{(j)}(t) - F^{(j+1)}(t) \right], & t < T\\
0, & t \geq T.
\end{cases}
\end{equation*}
Thus the mean time to replacement is given by
\begin{equation}\label{eq3.19}
\int_{0}^{\infty} P\left[ S > t \right] dt = \int_{0}^{T}\bar{F}(t)dt  + \sum_{j=1}^{N-1} G^{(j)}(Z) \int_{0}^{T} \left[F^{(j)}(t) - F^{(j+1)}(t)\right] dt.
\end{equation}
Then, dividing the expected cost for replacement (\ref{eq3.18}) by the mean time to replacement (\ref{eq3.19}), we get the expected cost rate (\ref{eq3.20}). 

\section*{\small Acknowledgement}
Prajamitra Bhuyan is supported by Research Impulse Award (Reference: DRI122PB), Department of Mathematics, Imperial College London.

\bibliographystyle{apalike}
\bibliography{bbl_main}

\begin{thebibliography}{}

\bibitem[Bhuyan and Dewanji, 2017a]{Bhuyan2017b}
Bhuyan, P. and Dewanji, A. (2017a).
\newblock Estimation of reliability with cumulative stress and strength
  degradation.
\newblock {\em Statistics -- A Journal of Theoretical and Applied Statistics},
  51(4):766--781.

\bibitem[Bhuyan and Dewanji, 2017b]{Bhuyan2017a}
Bhuyan, P. and Dewanji, A. (2017b).
\newblock Reliability computation under dynamic stress--strength modeling with
  cumulative stress and strength degradation.
\newblock {\em Communications in Statistics -- Simulation and Computation},
  46(4):2701--2713.

\bibitem[Chikte and Deshmukh, 1981]{Chikte1977}
Chikte, S. and Deshmukh, S. (1981).
\newblock Preventive maintenance and replacement under additive damage.
\newblock {\em Naval Research Logostics Quarterly}, 28(1):33--46.

\bibitem[Dowsland, 1995]{Dowsland1995}
Dowsland, K. (1995).
\newblock {\em Simulated annealing in modern heuristic techniques for
  combinatorial problems}.
\newblock McGraw-Hill.

\bibitem[Ebrahimi and Ramallingam, 1993]{Ebrahimi1993}
Ebrahimi, N. and Ramallingam, T. (1993).
\newblock Estimation of system reliability in brownian stress-strength models
  based on sample paths.
\newblock {\em Annals of the Institute of Statistical Mathematics}, 45.

\bibitem[Endharta and Yun, 2014]{Endharta2014}
Endharta, A. and Yun, W. (2014).
\newblock A comparison study of replacement policies for a cumulative damage
  model.
\newblock {\em International Journal of Reliability, Quality and Safety
  Engineering}, 21(4).

\bibitem[Eryilmaz, 2017]{Ery}
Eryilmaz, S. (2017).
\newblock Computing optimal replacement time and mean residual life in
  reliability shock models.
\newblock {\em Computers \& Industrial Engineering}, 103:40--45.

\bibitem[Kirkpatrick et~al., 1983]{Kirkpatrick1983}
Kirkpatrick, S., Gelatt, C., and Vecchi, M. (1983).
\newblock Optimization by simulated annealing.
\newblock {\em Science}, 220(4598):671--680.

\bibitem[Kotz et~al., 2000]{Kotz2000}
Kotz, S., Balakrishnan, N., and Johnson, N.~L. (2000).
\newblock {\em Continuous multivariate distributions : Models and
  applications}.
\newblock John Wiey and Sons Inc.

\bibitem[Lee and Cha, 2018]{Lee}
Lee, H. and Cha, J.~H. (2018).
\newblock A bivariate optimal replacement policy for a system subject to a
  generalized failure and repair process.
\newblock {\em Applied Stochastic Models in Business and Industry}.

\bibitem[Nakagawa, 1976]{Nakagawa1976b}
Nakagawa, T. (1976).
\newblock On a replacement problem of a cumulative damage model.
\newblock {\em Operations Research Quaterly}, 27(4):895--900.

\bibitem[Nakagawa, 2005]{Nakagawa2005}
Nakagawa, T. (2005).
\newblock {\em Maintenance Theory of Reliability}.
\newblock Springer--Verlag London Ltd.

\bibitem[Nakagawa, 2007]{Nakagawa2007}
Nakagawa, T. (2007).
\newblock {\em Shock and damage models in reliability theory}.
\newblock Springer -- Verlag London Ltd.

\bibitem[Piessens et~al., 1983]{Piessens1983}
Piessens, R., Doncker-Kapenga, E.~D., Uberhuber, C., and Kahaner, D. (1983).
\newblock {\em Quadpack : a subroutine package for automatic integration}.
\newblock Springer-Verlag.

\bibitem[Rafiee et~al., 2015]{Rafiee}
Rafiee, K., Feng, Q., and Coit, D.~W. (2015).
\newblock Condition-based maintenance for repairable deteriorating systems
  subject to a generalized mixed shock model.
\newblock {\em IEEE Transactions on Reliability}, 64(4):1164--1174.

\bibitem[Satow et~al., 2000]{Satow2000}
Satow, T., Teramoto, K., and Nakagawa, T. (2000).
\newblock Optimal replacement policy for a cumulative damage model with time
  deterioration.
\newblock {\em Mathematical and Computer Modelling}, 31:313--319.

\bibitem[Scarf et~al., 1996]{Scarf1996}
Scarf, P., Wang, W., and Laycock, P. (1996).
\newblock A stochastic model of crack growth under periodic inspection.
\newblock {\em Reliability Engineering \& System Safety}, 51:331--339.

\bibitem[Shue et~al., 2019]{Sheu}
Shue, S.~H., Liu, T.~H., Tsai, H.~N., and Zhang, Z.~G. (2019).
\newblock Optimization issues in k-out-of-n systems.
\newblock {\em Applied Mathematical Modelling}, 73:563--580.

\bibitem[Sobczyk, 1987]{Sobczyk1987}
Sobczyk, K. (1987).
\newblock Stochastic models for fatigue damage of materials.
\newblock {\em Advances in Applied Probability}, 19:652--673.

\bibitem[Sobczyk and Spencer, 1992]{Sobczyk1992}
Sobczyk, K. and Spencer, B. (1992).
\newblock {\em Random fatigue : From Data to Theory}.
\newblock Academic Press, New York.

\bibitem[Taylor, 1975]{Taylor1975}
Taylor, H. (1975).
\newblock Optimal replacement under additive damage and other failure models.
\newblock {\em Naval Research Logistics Quarterly}, 22(1):1--18.

\bibitem[Zhao et~al., 2018]{Zhao}
Zhao, X., Cai, K., Wang, X., and Song, Y. (2018).
\newblock Optimal replacement policies for a shock model with a change point.
\newblock {\em Computers \& Industrial Engineering}, 118:383--393.

\bibitem[Zhao et~al., 2013]{Zhou2013}
Zhao, X., Qian, C., and Nakagawa, T. (2013).
\newblock Optimal policies for cumulative damage models with maintenance last
  and first.
\newblock {\em Reliability Engineering and System Safety}, 110.

\bibitem[Zhou et~al., 2016]{Zhou2016}
Zhou, X., Wu, C., Li, Y., and Xi, L. (2016).
\newblock A preventive maintenance model for leased equipment subject to
  internal degradation and external shock damage.
\newblock {\em Reliability Engineering and System Safety}, 154.

\bibitem[Zuckerman, 1977]{Zuckerman1977}
Zuckerman, D. (1977).
\newblock Replacement models under additive damage.
\newblock {\em Naval Research Logistics Quarterly}, 24(4):549--558.

\end{thebibliography}
	
\end{document}